\documentclass[floatfix,noshowpacs,nofootinbib]{revtex4}%
\usepackage{graphicx}
\usepackage{subfigure}
\usepackage{amsmath,amssymb,amsfonts}
\usepackage{graphicx}
\usepackage{epsfig}
\usepackage{dcolumn}% Align table columns on decimal point
\usepackage{booktabs,threeparttable}% insert table package
\usepackage{multirow,array}
\usepackage{makecell}

\begin{document}

\title{Metal-insulator transition in correlated two-dimensional systems with disorder}
\author{Dragana Popovi\'c}
\affiliation{National High Magnetic Field Laboratory, Florida State University, Tallahassee, Florida 32310, USA}
\vspace{12pt}
\vspace{12pt}

\begin{abstract}
Experimental evidence for the possible universality classes of the metal-insulator transition (MIT) in two dimensions (2D) is discussed.  Sufficiently strong disorder, in particular, changes the nature of the transition.  Comprehensive studies of the charge dynamics are also reviewed, describing evidence that the MIT in a 2D electron system in silicon should be viewed as the melting of the Coulomb glass.  Comparisons are made to recent results on novel 2D materials and quasi-2D strongly correlated systems, such as cuprates.
\end{abstract}
\maketitle

\section{2D Metal-Insulator Transition as a Quantum Phase Transition}
\label{sec:intro}

\begin{figure}[t]
\vspace*{-0.2in}
\centerline{\includegraphics[width=4.5in]{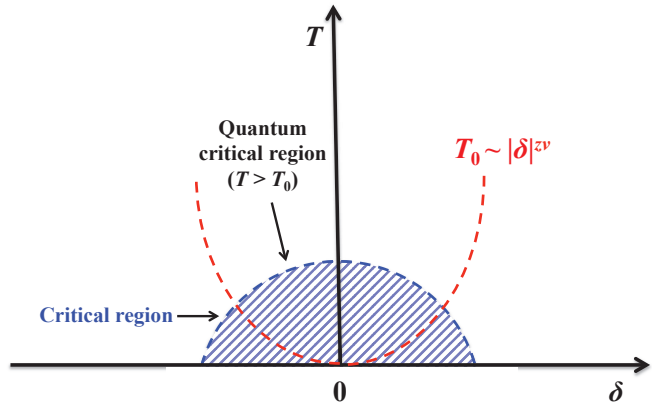}}
\vspace*{-0.75in}\caption{\label{fig:QPT} The phase diagram of a system near a quantum phase transition.  A quantum critical point at $T=0$ and $\delta=0$ separates two ground states, \textit{e.g.} an insulator and a metal.  The red dashed line denotes a smooth crossover at $T_{0}\sim |\delta|^{\nu z}$.  Well-defined differences between the two phases, such as insulating vs. metallic behavior, are observed only at $T<T_0$.  At $T>T_0$, the system is in the ``quantum critical region'', in which both quantum and thermal fluctuations are important.  Scaling behavior is expected in the entire (blue hatched) critical region.}
\end{figure}

The metal-insulator transition (MIT) in 2D systems remains one of the most fundamental open problems in condensed matter physics \cite{2DMIT-review_2001, 2DMIT-review_2004, 2DMIT-review_2010, Vlad-CIQPT}.  The very existence of the metal and the MIT in 2D had been questioned for many years but, recently, considerable experimental evidence has become available in favor of such a transition.  Indeed, in the presence of electron-electron interactions, the existence of the 2D MIT does not contradict any general idea or principle (see also chapters by V. Dobrosavljevi\'c, and A. A. Shashkin and S. V. Kravchenko).  It is important to recall that a qualitative distinction between a metal and an insulator exists only at temperature $T=0$: the conductivity $\sigma(T=0)\neq 0$ in the metal, and $\sigma(T=0)=0$ in the insulator.  Therefore, the MIT is an example of a quantum phase transition (QPT) \cite{Sachdev-book}: it s a continuous phase transition that occurs at $T = 0$, \textit{i.e.} between two ground states.  It is controlled by some parameter of the Hamiltonian of the system, such as carrier density, the external magnetic field, or pressure, and quantum fluctuations dominate the critical behavior.  In analogy to thermal phase transitions, a QPT is characterized by a correlation length $\xi\propto |\delta|^{-\nu}$ and the corresponding timescale $\tau_{\xi}\sim\xi^{z}$, both of which diverge in a power-law fashion at the critical point.  Here $\delta$ is a dimensionless (reduced) distance of a control parameter from its critical value, $\nu$ is the correlation length exponent and $z$ is the dynamical exponent.  Because of the Heisenberg uncertainty principle, in a QPT there is a characteristic energy (temperature) scale 
\begin{equation}
T_{0}\sim \frac{\hbar}{\tau_{\xi}}\sim |\delta|^{\nu z},
\end{equation}
which decreases to zero in a power-law fashion as the critical point is approached (Fig.~\ref{fig:QPT}).  

The existence of a diverging correlation length
leads to the expectation that, in the vicinity of a critical point, the behavior of the system will exhibit a certain degree of universality, independent of microscopic details.  Indeed, while microscopic theories describing the MIT remain controversial, scaling behavior is a much more robust and general property of second-order phase transitions (\textit{e.g.} \cite{Vlad-CIQPT}).  In the case of a MIT, which is controlled by changing the carrier density $n$, it is the conductivity that can be described in the critical region (Fig.~\ref{fig:QPT}) by a scale-invariant form as 
\begin{equation}
\label{eq:scaling}
\sigma(n, T)=\sigma_c(T)f(T/T_0)=\sigma_c(T)f(T/|\delta|^{z\nu}),
\end{equation}
where $\delta\equiv\delta_n=(n-n_c)/n_c$ ($n_c$ is the critical density) and 
\begin{equation}
\label{eq:x}
\sigma_c(T)\equiv\sigma(n=n_c, T)\propto T^{x}
\end{equation}
is the critical conductivity.  From the scaling description, it also follows that, in the metallic phase at $T=0$, 
\begin{equation}
\label{eq:mu}
\sigma(n,T=0)\propto\delta_{n}^{\mu},
\end{equation}
such that the conductivity exponent $\mu=x(z\nu)$.  Thus the behavior observed in the $T=0$ limit [Eq.~(\ref{eq:mu})] also provides a stringent test of the scaling behavior observed at $T\neq 0$ [Eqs.~(\ref{eq:scaling}) and (\ref{eq:x})].  One of the goals of the scaling analysis is to reveal critical exponents that characterize the transition and correspond to specific universality classes.  

The simplest scaling scenario predicts the validity of ``Wegner scaling'' for which the conductivity exponent $\mu=(D-2)\nu$, where $D$ is the dimensionality of the system.  This would imply that, at the 2D MIT, $x=(D-2)/z=0$, so the critical conductivity should not depend on $T$.  It is for this reason that many early studies of the 2D MIT identified $n_c$ as the carrier density where $d\sigma/dT$ changes sign from insulator-like ($d\sigma/dT>0$) to metallic ($d\sigma/dT<0$).  However, it should be emphasized that scaling with $x\neq 0$ for 2D systems does not contradict any fundamental principle \cite{Belitz}.  Indeed, such violations of ``Wegner scaling'' were predicted in the presence of ``dangerously irrelevant operators'' \cite{Amit}, for example for certain microscopic models with strong spin-dependent components of the Coulomb interactions \cite{noWegner1, noWegner2, noWegner3}.  As discussed below, it is precisely the general scaling form, Eq.~(\ref{eq:scaling}) with $x\neq 0$, that provides a satisfactory and consistent description of all the data near a 2D MIT. 

In general, a QPT can affect the behavior of the system up to surprisingly high temperatures. In fact, many unusual properties of various strongly correlated materials have been attributed to the proximity of quantum critical points. An experimental signature of a QPT at nonzero $T$ is the observation of scaling behavior with relevant parameters in describing the data.  At $T\neq 0$, however, the correlation length is finite, so single-parameter scaling in Eq.~(\ref{eq:scaling}) will work only if the sample size $L>\xi$.  Otherwise, the scaling function will depend not only on $T/T_0$, but also on another scaling variable $\sim L/\xi$.  Such finite-size effects have not been observed in the experimental studies of the 2D MIT so far: the scaling functions have been found to depend only on $T/T_0$, indicating that sample sizes have been sufficiently large for the experimental temperature range.

In real systems near a MIT, both disorder and electron-electron interactions may play an important role and lead to out-of-equilibrium or glassy behavior of electrons (\textit{e.g.} \cite{Dragana-CIQPT}).  The goal of this chapter is thus twofold.  First, it will focus on studies of the critical behavior of conductivity in 2D electron systems with different amounts and types of disorder, as well as different ranges of Coulomb interactions, in order to identify possible universality classes of the 2D MIT.  Second, experimental results obtained using a variety of protocols to probe charge dynamics will be reviewed, providing important information about the nature of the insulating phase and the 2D MIT.  The focus will be on detailed and comprehensive studies that have been carried out so far only on a 2D electron system (2DES) in (100)-Si metal-oxide-semiconductor field-effect transistors (MOSFETs).  However, recent results on novel 2D systems, in particular, a single- or few-layer transition metal dichalcogenides, and on quasi-2D strongly correlated materials, such as cuprates, will be also discussed.

\section{Critical Behavior of Conductivity}
\label{sec:critical}

2DESs in semiconductor heterostructures \cite{AFS} are  relatively simple systems for exploring the interplay of electronic correlations and disorder, because all the relevant parameters -- carrier density $n_s$, disorder and interactions -- can be varied easily.  For example, $n_s$ can be tuned over two orders of magnitude by applying voltage $V_g$ to the gate electrode.  At low $n_s$, the 2DES is strongly correlated: $r_s\gg 1$, where $r_s=E_{C}/E_{F}\propto n_{s}^{-1/2}$ ($E_C$ is the average Coulomb energy per electron and $E_F$ is the Fermi energy; see also \cite{2DMIT-review_2004} for more details).  Therefore, the effects of interactions become increasingly important as $n_s$ is reduced.

In Si MOSFETs, random potential (disorder) that is ``felt'' by the 2DES is caused by charged impurities (Na$^{+}$ ions), which are randomly distributed in the oxide and thus spatially separated from the 2DES.  In these devices, the (Drude) mobility $\mu=\sigma/(n_{s}e)$ of the 2DES peaks as a function of $n_s$ because, at very high $n_s$ that are not of interest here, the scattering due to the roughness of the Si-SiO$_2$ interface becomes dominant  \cite{AFS}.  The peak mobility at 4.2~K, $\mu_{\textrm{peak}}$, is commonly used as a rough measure of the amount of disorder.  As described below, detailed studies of $\sigma(n_s,T)$ demonstrate scaling behavior consistent with the existence of a QPT in all 2DESs in Si regardless of the amount of disorder.

\subsection{Role of disorder}
\label{disorder}

\subsubsection{Low-disorder samples}
\label{sec:low}

Si MOSFETs with relatively little disorder ($\mu_{\textrm{peak}}$[m$^2$/Vs]$\sim 1-3$) make it possible to access lower electron densities, and thus the regime of stronger interactions, before strong localization sets in.\footnote{In the strongly localized or insulating regime, the conductivity decreases exponentially with decreasing temperature, leading to $\sigma(T=0)=0$.}  The corresponding values of $r_s$ near the critical density have been $r_s> 13$.  In such low-disorder samples, the most obvious feature that suggests the existence of a metal is the large, almost an order-of-magnitude increase of conductivity with decreasing temperature ($d\sigma/dT <0$) observed at\footnote{The Fermi temperature $T_F$[K]$=7.31 n_s[10^{11}$cm$^{-2}]$ for electrons in Si MOSFETs \cite{AFS}.}  $T<T_F$ and low $n_s$, where $r_s\gg 1$.  However, this does \textit{not} necessarily imply $\sigma(T=0)\neq 0$, i.e. a metallic ground state.  Therefore, even though such a strong $d\sigma/dT <0$ had been known for a long time (e.g. \cite{Peter-Smith}), it was not considered as evidence for a metallic state.  Indeed, it was only after dynamical scaling (\ref{eq:scaling}) was demonstrated \cite{Krav-scal2} and independently confirmed on a different set of devices \cite{Popovic1997} that the problem of the 2D MIT attracted renewed attention.  

\begin{figure}[t]
\centerline{\includegraphics[width=3.6in]{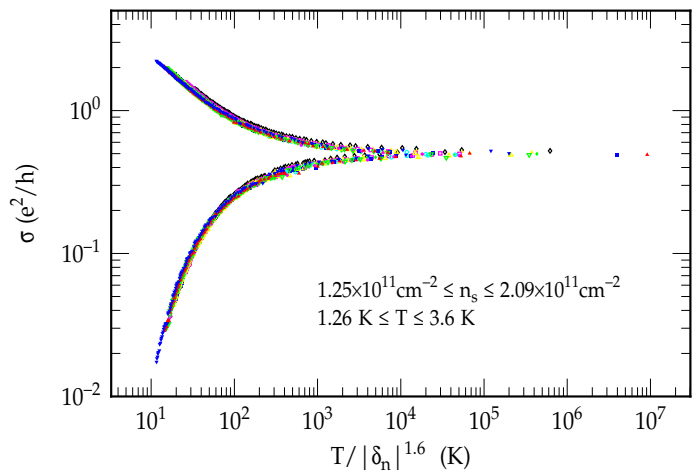}}
\caption{\label{fig:scaling} Scaling of conductivity in a low-disorder Si MOSFET with $\mu_{\textrm{peak}}\approx 1$~m$^2$/Vs (from \cite{Popovic1997}).  The scaling ranges of $n_s$ and $T$ are given on the plot;  $n_c=(1.67\pm 0.02)\times 10^{11}$cm$^{-2}$.}
\end{figure}

In those experiments, it was possible to collapse all the $\sigma(n_s, T)$ data near $n_c$ onto the same function $f(T/T_0)$ with two branches, the upper one for the metallic side of the transition and the lower one for the insulating side (see Fig.~\ref{fig:scaling}), using the same values $z\nu=1.6$ and $x=0$.  The 
temperature ranges for scaling were also comparable: $0.08 \lesssim T/T_{F} \lesssim 0.4$ in Fig.~\ref{fig:scaling} \cite{Popovic1997} and $0.05 \lesssim T/T_{F} \lesssim 0.3$ in Ref.~\cite{Krav-scal2}.  The critical density $n_c$ was identified as the ``separatrix'', i.e. the density $n_{s}^{\ast}$ where $d\sigma/dT$ changes sign.  Subsequently, other, more appropriate methods were used to identify $n_c$.  In particular, $n_c$ was determined based on both a vanishing activation energy \cite{Shashkin-nc, JJ_PRL02, Jan2004} and a vanishing nonlinearity of current-voltage ($I-V$) characteristics when extrapolated from the insulating phase \cite{nc1,Shashkin-nc}, and established that indeed $n_c\approx n_{s}^{\ast}$.  It is important to stress, however, that this is true only in low-disorder samples with non-magnetic scattering.  The situation is very different in the presence of scattering by local magnetic moments and in highly disordered samples.  In those cases, the two densities can be vastly different, as shown in Sections~\ref{sec:moments} and \ref{sec:high}.

\begin{figure}
\includegraphics[width=3.2in]{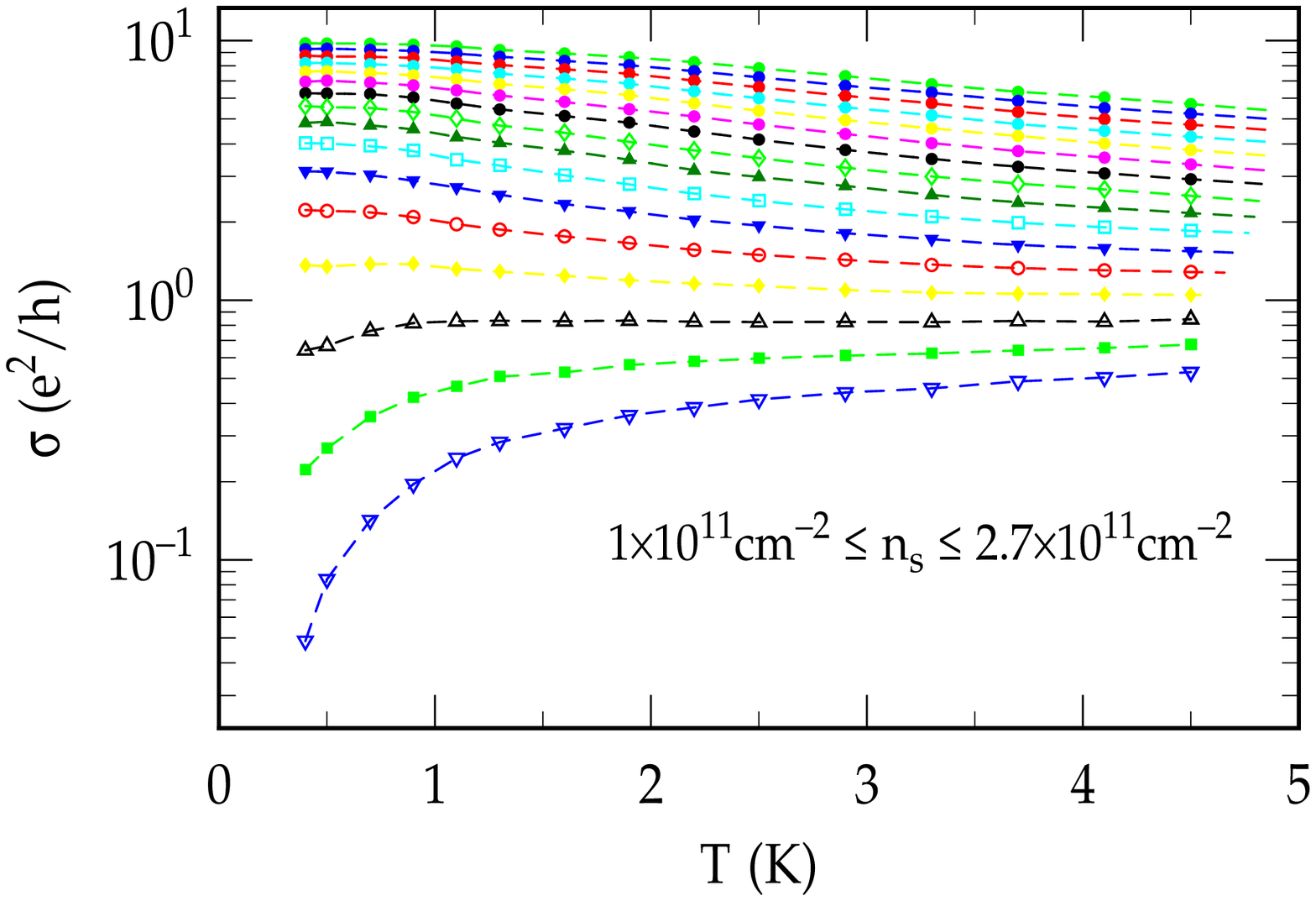}
\includegraphics[width=3.2in]{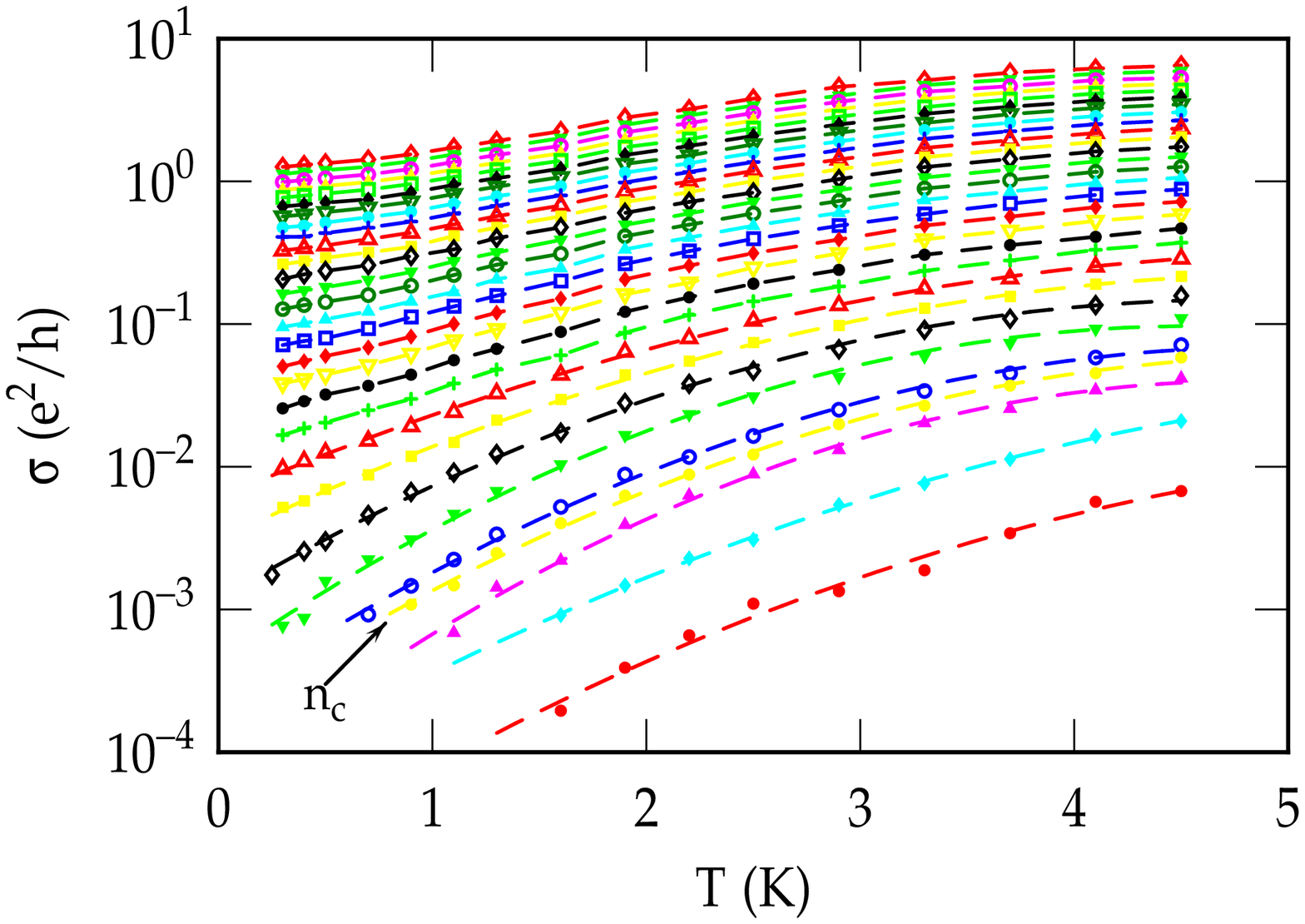}
\caption{\label{fig:sigmaT} Conductivity $\sigma$ vs $T$ for different $n_s$ in a low-disorder 2DES with $\mu_{\textrm{peak}}\approx 0.5-1$~m$^2$/Vs.  Left: The substrate (back-gate) bias on the sample was $V_{sub}=-40$~V (adapted from \cite{Feng1999}).  The data can be scaled with $x=0$ (similarly to Fig.~\ref{fig:scaling}) around the ``separatrix'' (open black triangles) $n_{s}^{\ast}=1.25\times 10^{11}$cm$^{-2}$ only at $T> 1$~K.  Right: $V_{sub}=+1$~V, which induces scattering by local magnetic moments; $0.3\leq n_s(10^{11}$cm$^{-2})\leq 3.0$ (from \cite{Feng2001}).  The data can be scaled (Fig.~\ref{fig:T2scaling}) around $n_c=0.55\times 10^{11}$cm$^{-2}$ with $x\neq 0$ down to the lowest $T$.}
\end{figure}

Therefore, the first studies of scaling near a 2D MIT seemed to be consistent with ``Wegner scaling'', in which the critical conductivity (\ref{eq:x}) does not depend on temperature, i.e. the exponent $x=0$.  Not long after, it was realized that this simple scaling scenario fails in the metallic regime at the lowest temperatures where, following its initial large increase, $\sigma(T)$ becomes a weak function as $T\rightarrow 0$ \cite{Pudalov1998,nc4}.  A careful analysis of the data showed \cite{njk2,njk3} that they could be scaled according to the more general form (\ref{eq:scaling}) with $\sigma_c=\sigma(n_c,T)\propto T^x$, $x\neq 0$ (the scaling range was $0.02 \lesssim T/T_{F} \lesssim 0.2$).  This type of scaling implies, of course, that $\sigma_c$ vanishes as $T\rightarrow 0$, i.e. $n_c$ does not coincide with the ``separatrix'' $n_{s}^{\ast}$, but instead $\sigma_c(T)$ belongs to the insulating family of curves. This conclusion is consistent with the results of activation energy and nonlinear $I-V$ studies that reported a small but systematic difference of a few percent such that $n_c<n_{s}^{\ast}$ \cite{Pudalov1998,nc4}.  Nevertheless, it has proven difficult to determine the exponent $x$ reliably, because the weak, low-$T$ behavior of $\sigma(T)$ is observable only over a narrow range of very low temperatures ($T\sim 0.1$~K), when it becomes difficult to cool the electrons in the experiment.  In samples that are slightly more disordered ($\mu_{\textrm{peak}}\lesssim 1$~m$^2$/Vs), the onset of the weak $\sigma(T)$ (or an apparent ``saturation'' on a logarithmic scale) in the metallic regime takes place at higher temperatures, e.g. at $T\lesssim 1$~K in Fig.~\ref{fig:sigmaT}(left), where it is easy to cool the electrons.
This shows that the weak metallic $\sigma(T)$  (``saturation'') observed at low $T$ is an intrinsic effect and its onset depends on disorder.  Here it is still possible to scale the data at $T> 1$~K with $z\nu=1.6$ and $x=0$ around the ``separatrix'' (open black triangles in Fig.~\ref{fig:sigmaT}(left) corresponding to $n_{s}^{\ast}=1.25\times 10^{11}$cm$^{-2}$), but it is obvious that this curve acquires an insulating $T$-dependence at $T<1$~K, so that ``Wegner scaling'' will fail.  However, it is still difficult to make reliable fits to $\sigma(T)$ in the low-$T$ regime, so that the direct determination of the exponent $x$ in low-disorder samples remains a challenge.

On the other hand, by extrapolating the metallic ``saturation'' of $\sigma$ to $T=0$, it has been found \cite{Fletcher2001} that $\sigma(n_s,T=0)\propto\delta_{n}^{\mu}$, with the exponent $\mu=1-1.5$.  The critical density $n_c$ obtained in this way was by about 1-5\% lower than $n_{s}^{\ast}$, in agreement with the results of the activation energy and nonlinear $I-V$ studies.  Some estimates of the exponent $x$ may be made then based on the relation $\mu=x(z\nu)$.

Scaling with $x\neq 0$ is also observed in 3D materials near the MIT \cite{Myriam_review}.  Other striking similarities near the MIT between transport properties of a 2DES in low-disorder Si MOSFETs and those observed in Si:B, a 3D system, were pointed out early on \cite{Popovic1997}.  For example, in Si:B, magnetoconductance, which is negative, depends strongly on the carrier concentration and shows a dramatic decrease at the transition \cite{3Dbeta-2}.  Such anomalous behavior was attributed to electron-electron interactions.  In a 2DES, the negative component of the magnetoconductance, which arises from the spin-dependent part of the electron-electron interaction \cite{LR}, also decreases rapidly at the transition, suggesting that electron-electron interactions may play a similar role in both 2D and 3D systems near the MIT \cite{Popovic1997}.

In recent years, many different types of experiments, including both transport and thermodynamic measurements, have been performed on low-disorder samples.  Some of those studies are discussed in detail in the chapter by A. A. Shashkin and S. V. Kravchenko.  There is mounting evidence that suggests that electron-electron interactions are responsible for a variety of phenomena observed in the metallic regime of low-disorder 2DES near the MIT, including a large increase of  $\sigma$ with decreasing temperature ($d\sigma/dT<0$) \cite{Radonjic}.  In particular, since the most striking experimental features are not sensitive to weak disorder (see, e.g., the thermopower study \cite{Kravchenko-thermo}),  they have been interpreted as evidence that the MIT in such low-disorder systems is driven by electron-electron interactions and that disorder has only a minor effect.  It is thus important to compare the properties of the MIT, such as critical exponents, in low-disorder samples to those obtained in samples with a different type or amount of disorder.

\subsubsection{Special disorder: local magnetic moments}
\label{sec:moments}

It turns out that the metallic behavior with $d\sigma/dT<0$ is easily suppressed by scattering of the conduction electrons by disorder-induced local magnetic moments.  More precisely, even an arbitrarily small amount of such scattering is sufficient to suppress the $d\sigma/dT<0$ behavior in the $T\rightarrow 0$ limit \cite{Feng1999}.  It is important to note that this does not necessarily indicate the destruction of the metallic phase.  In disordered 3D metals, for example, it is well known that the derivative $d\sigma/dT$ can be either negative or positive near the MIT \cite{LR}.  Indeed, as described below, in the presence of local magnetic moments, the 2DES exhibits in the metallic phase the simplest $\sigma(T)$, observed over two decades of $T$, allowing for an unambiguous extrapolation to $T=0$ and an excellent fit to the dynamical scaling described by Eq.~(\ref{eq:scaling}).

In Si MOSFETs, it is possible to change the disorder, for a fixed $n_s$, by applying bias $V_{sub}$ to the Si substrate (back gate) \cite{AFS}.  In particular, the reverse (negative) $V_{sub}$ moves the electrons closer to the interface, which increases the disorder.  It also increases the splitting between the subbands since the width of the triangular potential well at the interface in which the electrons are confined is reduced by applying negative $V_{sub}$.  Usually, only the lowest
subband is occupied at low $T$, giving rise to the 2D behavior.  However, in sufficiently disordered samples or wide enough potential wells, the band tails associated with the upper subbands can be so long that some of their strongly localized states may be populated even at low $n_s$, and act as additional scattering centers for 2D electrons.  In particular, since at least some of them must be singly populated due to a large on-site Coulomb repulsion (tens of meV), they may act as local magnetic moments.  Clearly, the negative $V_{sub}$ reduces this type of scattering by depopulating the upper subband.  Therefore, the effects of local magnetic moments on the transport properties of the 2DES were studied by varying $V_{sub}$ \cite{Feng1999,Feng2001}.  It is important to note that the presence of local moments did not have a significant effect on the value of $\mu_{\textrm{peak}}$, probably because the upper subband is depopulated at high $V_g$, where mobility peaks.\footnote{For a fixed $V_{sub}$, the subband splitting increases by increasing $V_g$ \cite{AFS}.}

\begin{figure}
\psfig{file=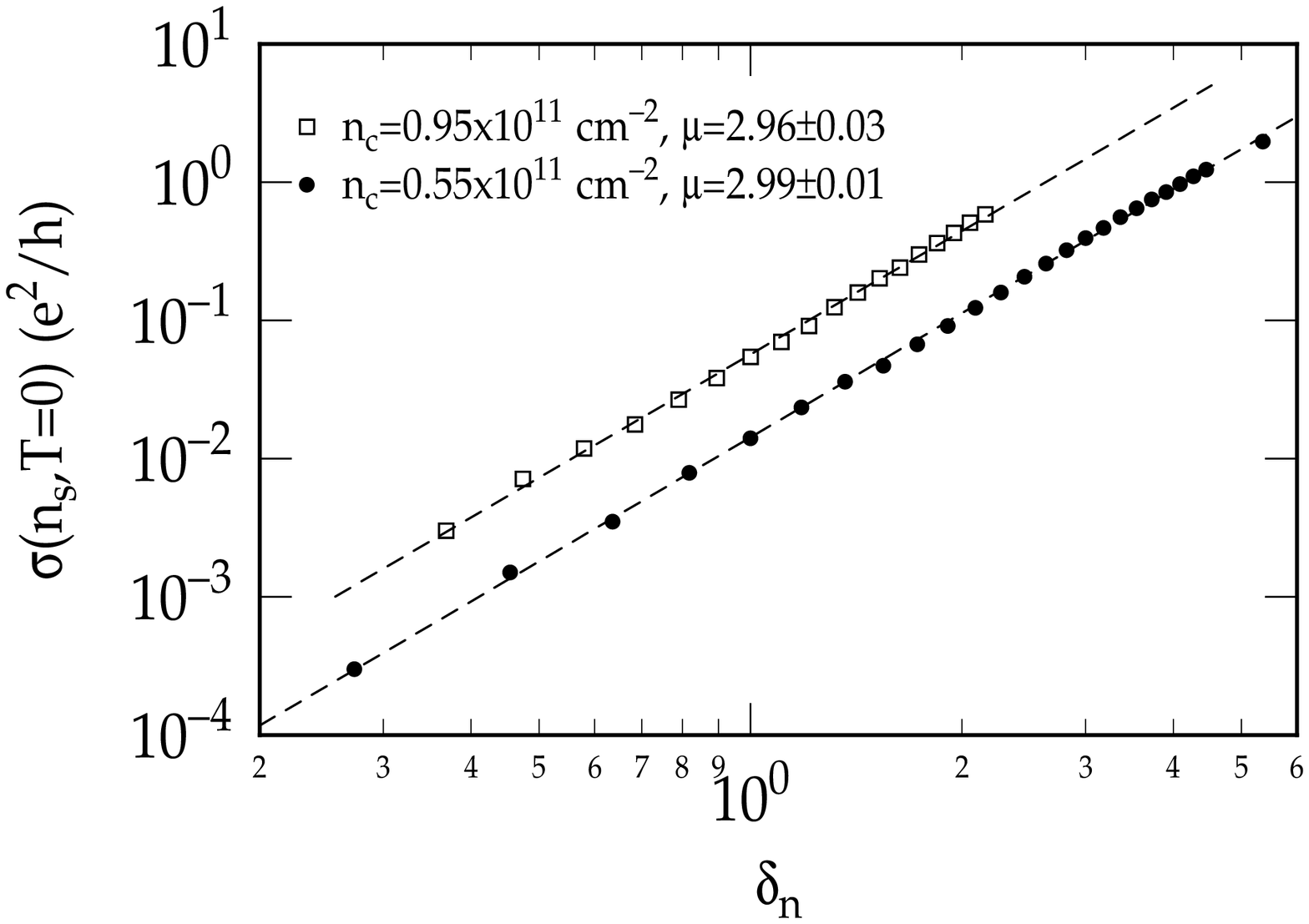,width=3.0in}
\psfig{file=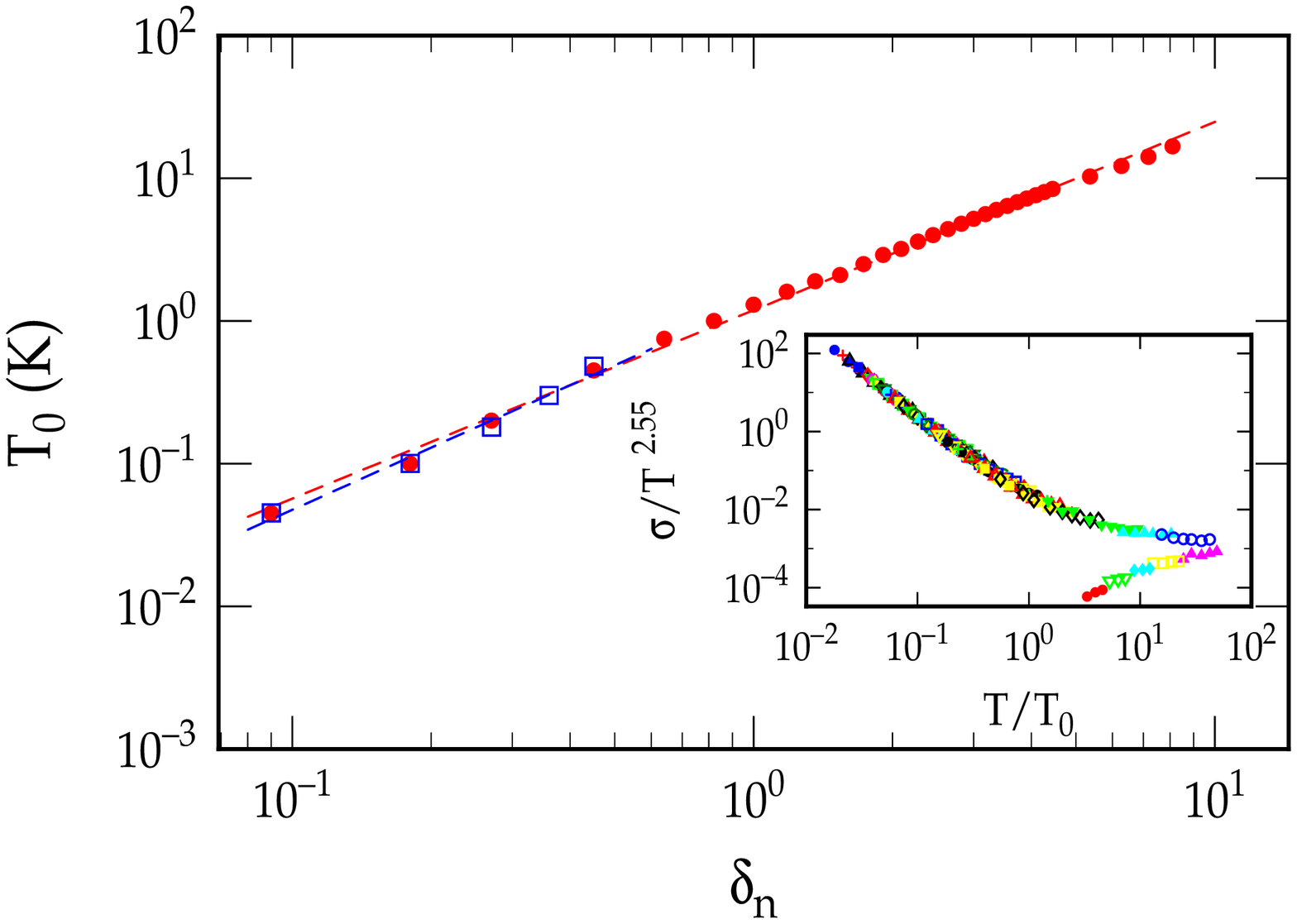,width=3.0in}
\caption{\label{fig:T2scaling} Left: $\sigma (n_s,T=0)$ vs $\delta_n$ for two low-disorder samples with different $n_c$, as shown, in the presence of scattering by local magnetic moments.  The dashed lines are fits with the slopes equal to the critical exponent $\mu$.  At the MIT, the corresponding $r_s\approx 17$ and 22 for the two samples, respectively.  Right: Scaling parameter $T_0$ as a function of $|\delta_n|$ for the sample with $n_c=0.55\times 10^{11}$cm$^{-2}$ [Fig.~\ref{fig:sigmaT}(right)]; open symbols: $n_s<n_c$, closed symbols: $n_s>n_c$.  The dashed lines are fits with slopes $z\nu=1.4\pm 0.1$ and $z\nu=1.32\pm 0.01$, respectively.  Inset: scaling of raw data $\sigma/\sigma_c\sim\sigma/T^x$ in units of $e^2/h$K$^{2.55}$ for all $n_s$ shown in Fig.~\ref{fig:sigmaT}(right) and $T<2$~K.  From \cite{Feng2001}.}
\end{figure}

Figure~\ref{fig:sigmaT} illustrates the effect of local magnetic moments on $\sigma(T)$ in the same sample: while $d\sigma/dT<0$ is observed at high $n_s$ in the absence of local moments (left panel), $\sigma(T)$ curves  become insulator-like ($d\sigma/dT>0$) for all $n_s$ after many local moments are introduced (right panel).  In the latter case, the critical density can be easily identified by plotting the data on a log-log scale (not shown): at $n_c$, the conductivity obeys a pure power-law form (\ref{eq:x}) with $x\approx 2.6$ \cite{Feng2001}.  For $n_s<n_c$, $\sigma(T)$ decreases exponentially, as expected in the insulating phase.  For $n_s>n_c$, a simple and precise form $\sigma(n_s,T)=\sigma(n_s,T=0)+A(n_s)T^2$ is observed from about 2~K down to the lowest accessible $T=0.020$~K \cite{Eng2002}.  Figure~\ref{fig:T2scaling}(left) shows that the extrapolated 
zero-temperature conductivity is a power-law function of $\delta_n$ (\ref{eq:mu}), i.e. $\sigma (n_s,T=0)\propto\delta_{n}^{\mu}$ ($\mu\approx 3$), as expected in the vicinity of a quantum critical point \cite{Goldenfeld}.  Furthermore, the data obey dynamical scaling (\ref{eq:scaling}) with $z\nu\approx 1.3$ [Fig.~\ref{fig:T2scaling}(right)] and a $T$-dependent critical conductivity (\ref{eq:x}) with $x\approx 2.6$, i.e. in agreement with theoretical expectations near a QPT.  The consistency of the scaling analysis is confirmed by comparing $\mu=x(z\nu)=3.4\pm 0.4$ obtained from scaling with $\mu=3.0\pm 0.1$ determined from the $T=0$ extrapolation of $\sigma(n_s,T)$ [Fig.~\ref{fig:T2scaling}(left)].

The exponent $z\nu$ is thus comparable to that found in low-disorder samples in the absence of scattering by local magnetic moments (Sec. \ref{sec:low}; see also Sec. \ref{sec:universal}).  On the other hand, unlike $z\nu$, the exponent $x$ seems to be more sensitive to the type of scattering (magnetic vs. nonmagnetic).  It is also noted that scaling with a $T$-dependent prefactor $\sigma_c(T)$ is not consistent with the simple single-parameter scaling hypothesis \cite{Vlad-scaling}, which allows only for a  $d\sigma/dT <0$ behavior in the conducting phase, but it does not violate any general principle and it is analogous to the MIT in 3D systems.

\subsubsection{High-disorder samples}
\label{sec:high}

The most interesting situation is found in high-disorder samples, where electron-electron interactions are still strong (e.g. $r_s\sim 7$ near $n_c$).  Indeed, the competition between disorder and interactions leads to the striking out-of-equilibrium or glassy behavior near the MIT and in the insulating regime.  The manifestations of glassiness in \verb"both" low- and high-disorder samples are discussed in Sec. \ref{sec:dynamics}.  However, charge dynamics has been studied in more detail in high-disorder 2DESs, and there is ample evidence for the MIT and for the importance of long-range Coulomb interactions also in these systems \cite{Dragana-CIQPT}.  Thus one of the key questions that arises is: What is the nature of the MIT in a high-disorder 2DES with interactions?  More precisely, is it dominated by disorder, or is it the same as the MIT in a low-disorder 2DES, which seems to be driven by interactions?  The scaling behavior discussed below provides compelling evidence (a) for the existence of a metal-insulator QPT and (b) that sufficiently strong disorder changes the universality class of the MIT.

\begin{figure}[t]
\vspace*{-0.495in}
\centerline{\includegraphics[width=3.7in,clip]{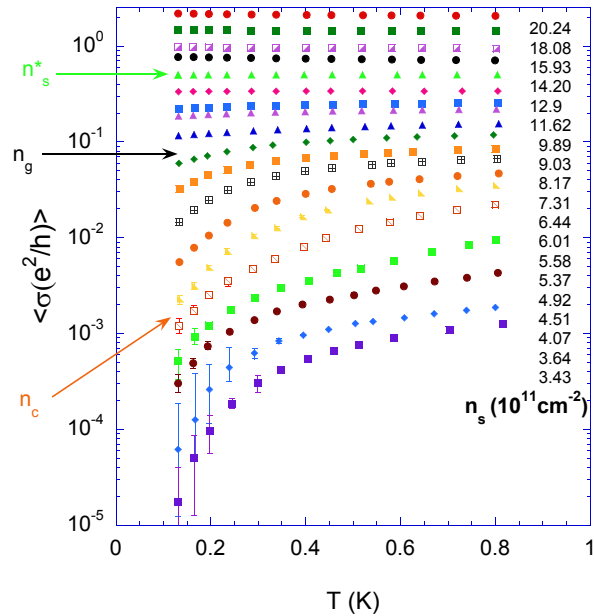}}
\vspace*{-0.8in}
\caption{\label{fig:high} Time-averaged conductivity vs $T$ for different $n_s$, as shown, in a high-disorder 2DES with $\mu_{\textrm{peak}}\approx 0.06$~m$^2$/Vs.  The arrows mark three densities: $n_{s}^{\ast}=12.9\times 10^{11}$cm$^{-2}$ is the ``separatrix'', $n_g=7.5\times 10^{11}$cm$^{-2}$ is the glass transition density, and $n_c=5.0\times 10^{11}$cm$^{-2}$ is the critical density for the MIT (the corresponding $r_s\sim 7$).  From \cite{Snezana2002}.}
\end{figure}

Detailed studies of transport and electron dynamics near the MIT have been performed on a set of Si MOSFETs\footnote{The substrate bias $V_{sub}=-2$~V was applied to maximize $\mu_\textrm{peak}$ by removing the contribution of scattering by local magnetic moments (see Sec.~\ref{sec:moments} above), at least in the experimental $T$-range.} with $\mu_\textrm{peak}\approx 0.06$~m$^2$/Vs.   A typical $\sigma(n_s,T)$ is shown in Fig.~\ref{fig:high}  Because of the glassy fluctuations of $\sigma$ with time $t$ at low $n_s$ and $T$, Fig.~\ref{fig:high} actually shows the time-averaged conductivity $\langle\sigma\rangle$, and the error bars correspond to the size of the fluctuations with time.  In this Section, which focuses on the behavior of the average conductivity, the notation $\sigma$ will be used instead of $\langle\sigma\rangle$ for simplicity.  It is important to note that scaling of the average conductance is expected to work even in the
presence of large fluctuations.  For example, it has been demonstrated that, for the case of Anderson transitions, scaling is, in fact, valid not only for the average conductance, but also for all moments even when relative fluctuations $\Delta\sigma/\langle\sigma\rangle\sim 1$ \cite{Evers}. Therefore, exploring the average conductance to get information about the MIT is well justified.

\begin{figure}
	\centering
		\includegraphics[width=9.0cm,clip]{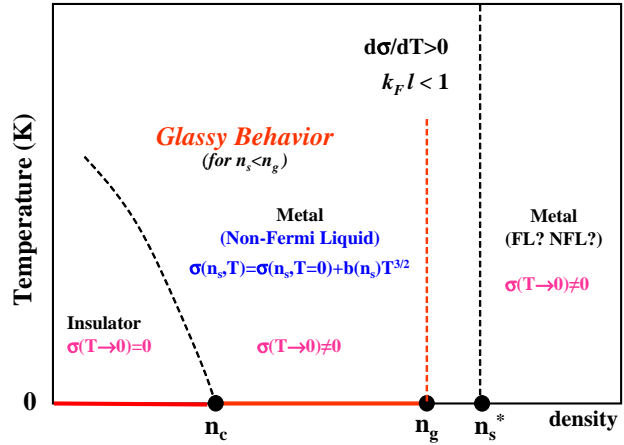}
\vspace*{-0.5cm}\caption{Experimental phase diagram of the 2DES in Si MOSFETs.  The MIT takes place at $n_c$ and $T=0$.  The nature of the metallic phase at high $n_s$, such that $k_{F}l>1$, is still under debate.  The glass transition occurs at $T_g=0$ for all $n_s<n_g$: the solid red line thus represents a line of quantum critical points.  In the intermediate, metallic glassy phase $n_c<n_s<n_g$, $\sigma(T)$ obeys a particular, non-Fermi liquid form.  The aging properties of the glass change abruptly at $n_c$, indicating different natures of the insulating and metallic glass phases.  In low-disorder samples, the intermediate phase vanishes: $n_c\lesssim n_{s}^{\ast}\approx n_g$.  From \cite{Dragana-CIQPT}.}
	\label{fig:summary}
\end{figure}

Similar to other high-disorder samples or samples with very low 4.2~K peak mobility ($\mu_\textrm{peak} < 0.1$~m$^2$/Vs), there is hardly any $d\sigma/dT<0$ metallic behavior observed at high $n_s$.  In analogy with studies of low-disorder 2DES (Sec. \ref{sec:low}), $n_{s}^{\ast}$ is defined as the density where $d\sigma/dT$ changes sign, and $n_c$ is determined by extrapolating the activation energies in the insulating regime to zero.  Surprisingly, here $n_c$ turns out to be more than a factor of two smaller than $n_{s}^{\ast}$ (Fig.~\ref{fig:high}).  For $n_s>n_c$, the low-$T$ data are best described by the metallic ($\sigma(T=0)>0$) power law $\sigma(n_s,T)=\sigma(n _s, T=0) + b(n_s)T^{1.5}$ \cite{Snezana2002}.  The surprising non-Fermi liquid (NFL) $T^{3/2}$ behavior is consistent with theory \cite{Dalidovich,Sachdev-2012} for the transition region between a Fermi liquid and an (insulating) electron glass.  Indeed (see Sec. \ref{sec:highdynamics}), the transition into a charge (Coulomb) glass in high-disorder samples takes place as $T\rightarrow 0$ at a density $n_g$, such that $n_c<n_g<n_{s}^{\ast}$ (Fig.~\ref{fig:high}).  The $T^{3/2}$ correction is characteristic of transport in the intermediate, $n_c < n_s < n_g$ region where the dynamics is glassy, but where $\sigma$ is still metallic [$\sigma(T\rightarrow 0)\neq 0$] albeit so small that $k_{F}l < 1$ (Fig. \ref{fig:summary}; $k_{F}$ -- Fermi wave vector, $l$ -- mean free path).  
For $k_{F}l < 1$, there is indeed no reason to expect standard Fermi liquid behavior.  The simple form of $\sigma(T)$ in the intermediate, metallic phase allows a reliable extrapolation to $T=0$.  The extrapolated values of $\sigma(n_s,T=0)$ go to zero at $n_c$ that is in agreement with that obtained from the data on the insulating side of the MIT, and $\sigma_c(T)\propto T^x$ with $x=1.5$.  For clarity, the main experimental observations are summarized in Fig.~\ref{fig:summary}.

\begin{figure}
\centering
\includegraphics[width=8.2cm]{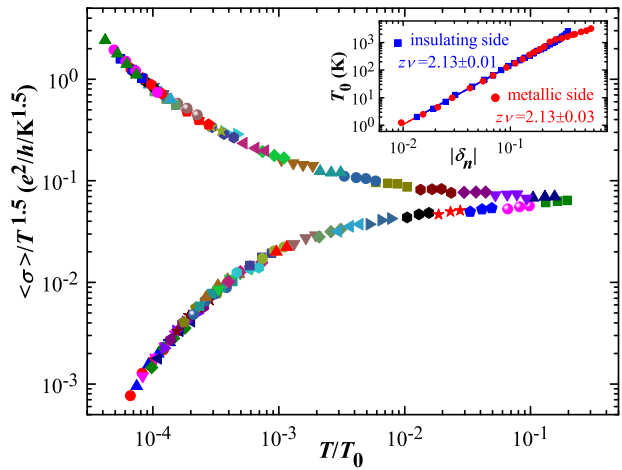}
\caption {Scaling of $\sigma/\sigma_{c}\propto\sigma/T^{x}$, $x=1.5$, for the high-disorder sample in Fig.~\ref{fig:high}  Different symbols correspond to $n_s$ from $3.45\times10^{11}$~cm$^{-2}$ to $8.17\times10^{11}$~cm$^{-2}$; $n_c= 5.22\times10^{11}$~cm$^2$.  It was possible to scale the data below $\sim 0.3$~K down to the lowest $T=0.13$~K (the lowest $T/T_F\approx 0.003$).  Inset: $T_0$ vs $\delta_n$.  The lines are fits with slopes $z\nu=2.13\pm0.01$ and $z\nu=2.13\pm0.03$ on the insulating and metallic sides, respectively.  From \cite{Ping2015}.}\label{fig:scalingthick}
\end{figure}

Figure~\ref{fig:scalingthick} demonstrates that, near $n_c$, the data exhibit dynamical scaling $\sigma(n_s, T)/\sigma_{c}(T)\propto\sigma(n_s, T)/T^{1.5}$, a signature of the QPT, also in this system.  The scaling parameter follows a power law, $T_0\propto|\delta_n|^{z\nu}$  (Fig.~\ref{fig:scalingthick} inset), in agreement with theoretical expectations near a QPT.  The value of the critical exponents $z\nu\approx2.1$ obtained here represents a major difference from the
%
%%%%%%%%%%%%%%%%%%%%%%
%%%%%%%%%%%%%%%%%%%%%%%%%%%%%%%%%%%%%%%%%%%%%%%%%%%%%%%%%%%%%
%
consistently lower $z\nu=1.0-1.7$  found in low-disorder 2DES \cite{Krav-scal2,Sarachik1996,Popovic1997, Pudalov1998, Feng1999, Feng2001, Ted-Belitz-2013}, indicating that sufficiently strong disorder changes the nature of the MIT from interaction-driven in low-disorder samples to disorder-dominated in high-disorder 2DES.  Before discussing different universality classes further (Sec.~\ref{sec:universal}), another central issue needs to be addressed first, namely, the role of the range of electron-electron interactions in the 2D MIT.

\subsection{Effects of the range of electron-electron interactions}
\label{interactions}

According to the scaling theory of localization \cite{gangof4}, all electrons in a disordered, noninteracting 2D system become localized at $T=0$.  Therefore, interactions must play a key role in stabilizing a metallic ground state in 2D.  An important question then is how the length scale of the Coulomb interactions controls the ground state and the properties of the MIT.  However, a vast majority of experiments on 2D systems (e.g. those discussed elsewhere in this chapter and in chapter by A. A. Shashkin and S. V. Kravchenko) have been carried out on devices in which Coulomb interactions are \verb"not" screened.  At the same time, the use of a nearby metallic gate or ground plane to limit the range of the Coulomb interactions between charge carriers in 2D systems is a well-known technique that has been explored both theoretically (see, e.g., \cite{Peeters1984,Widom1988,Hallam1996,Sushkov2009,Skinner-Shk,Skinner2010,Fregoso2013}) and experimentally, e.g., in the investigation of the melting of the Wigner crystal formed by electrons on a liquid He surface \cite{Mistura1997}. 

In the context of the 2D MIT in \verb"low-disorder "devices, screening by the gate has been used to explore the role of Coulomb interactions in the metallic \cite{Ho2008} and insulator-like \cite{Huang2014} regimes of a 2D hole system (2DHS) in AlGaAs/GaAs heterostructures  and in the metallic regime of Si MOSFETs \cite{Tracy}.  The focus of those studies, however, was on the form of $\sigma(T)$, and especially on the sign of $d\sigma/dT$.  In particular, there have been no studies of scaling and critical exponents, that is, of the properties of the quantum critical point itself.  

In the case of \verb"low-disorder" 2DES \verb"with local magnetic moments", there have been no studies at all of the effects of screened Coulomb interactions.

On the other hand, in \verb"high-disorder" Si MOSFETs, studies of scaling behavior of $\sigma(n_s, T)$ on both metallic and insulating sides of the MIT were carried out on devices in which the long-range part of the Coulomb interaction is screened by the gate \cite{Ping2015}.  The metallic gate at a distance $d$ from the 2DES creates an image charge for each electron, modifying the Coulomb interaction from $\sim1/r$ to $\sim[1/r-1/\sqrt{r^2+4d^2}]$.  When the mean carrier separation $a=(\pi n_s)^{1/2}\gg d$, this potential falls off in a dipole-like fashion, as $\sim 1/r^3$.  Therefore, in Si MOSFETs, the range of the electron-electron Coulomb interactions can be changed by varying the thickness of the oxide $d_{ox}=d$.  Importantly, measurements were done on a set of MOSFETs that had been fabricated simultaneously as the high-disorder devices with long-range Coulomb interactions discussed in Sec.~\ref{sec:high} (see also Figs. \ref{fig:high} and \ref{fig:scalingthick}).  In the latter case, $d_{ox}=50$~nm, comparable to that in other Si MOSFETs used in the vast majority of studies of the 2D MIT \cite{2DMIT-review_2001, 2DMIT-review_2004, 2DMIT-review_2010,Dragana-CIQPT}.  In the $d_{ox}=50$~nm samples, in the low-$n_s$ regime of interest near the MIT, the corresponding $5.3\lesssim d/a \leq 8.0$.  On the other hand, in ``thin-oxide'' devices with $d_{ox}=6.9$~nm \cite{Ping2015}, substantial screening by the gate is expected in the scaling regime of $n_s$ near the MIT, where $0.7\lesssim d/a \lesssim1.0$.  For comparison, in ground-plane screening studies carried out on low-disorder samples, $0.8\leq d/a \leq 1.8$ in Ref.~\cite{Tracy}, $1.1\lesssim d/a\leq 5$ in Ref.~\cite{Huang2014}, and $2\leq d/a\leq 19$ in Ref.~\cite{Ho2008}.

In general, the results obtained on high-disorder, thin-oxide samples resemble those on thick-oxide devices, as follows.\footnote{In analogy with Sec.~\ref{sec:high}, $\sigma$ is used here to denote the time-averaged conductivity $\langle\sigma\rangle$.}  

\begin{enumerate}

\item At the lowest $n_s$, $\sigma(T)$ decreases exponentially with decreasing $T$, in agreement with the 2D variable-range hopping law, indicating an insulating ground state.  The critical density $n_c$ is determined from the vanishing of the activation energy, as described earlier.

\item For $n_s>n_c$, the low-$T$ data are best described by the metallic power law $\sigma(n_s,T)=\sigma(n _s, T=0) + b(n_s)T^{1.5}$.  This type of non-Fermi-liquid behavior was found to be a characteristic of the intermediate, metallic glassy phase observed in high-disorder samples with long-range Coulomb interactions (Sec.~\ref{sec:high}), as well as in low-disorder 2DES in parallel magnetic field (Sec.~\ref{sec:field}).  At $n_c$, $\sigma_c\propto T^x$  with $x=1.5$.

\item On the metallic side of the MIT, the extrapolated $\sigma(n_s,T=0)$ go to zero [Fig.~\ref{fig:scalingthin}(b)] at the same density as $n_c$ 
%
%%%%%%%%%%%%%%%%%%%%%%
\begin{figure}
\centering
\includegraphics[width=7.8cm]{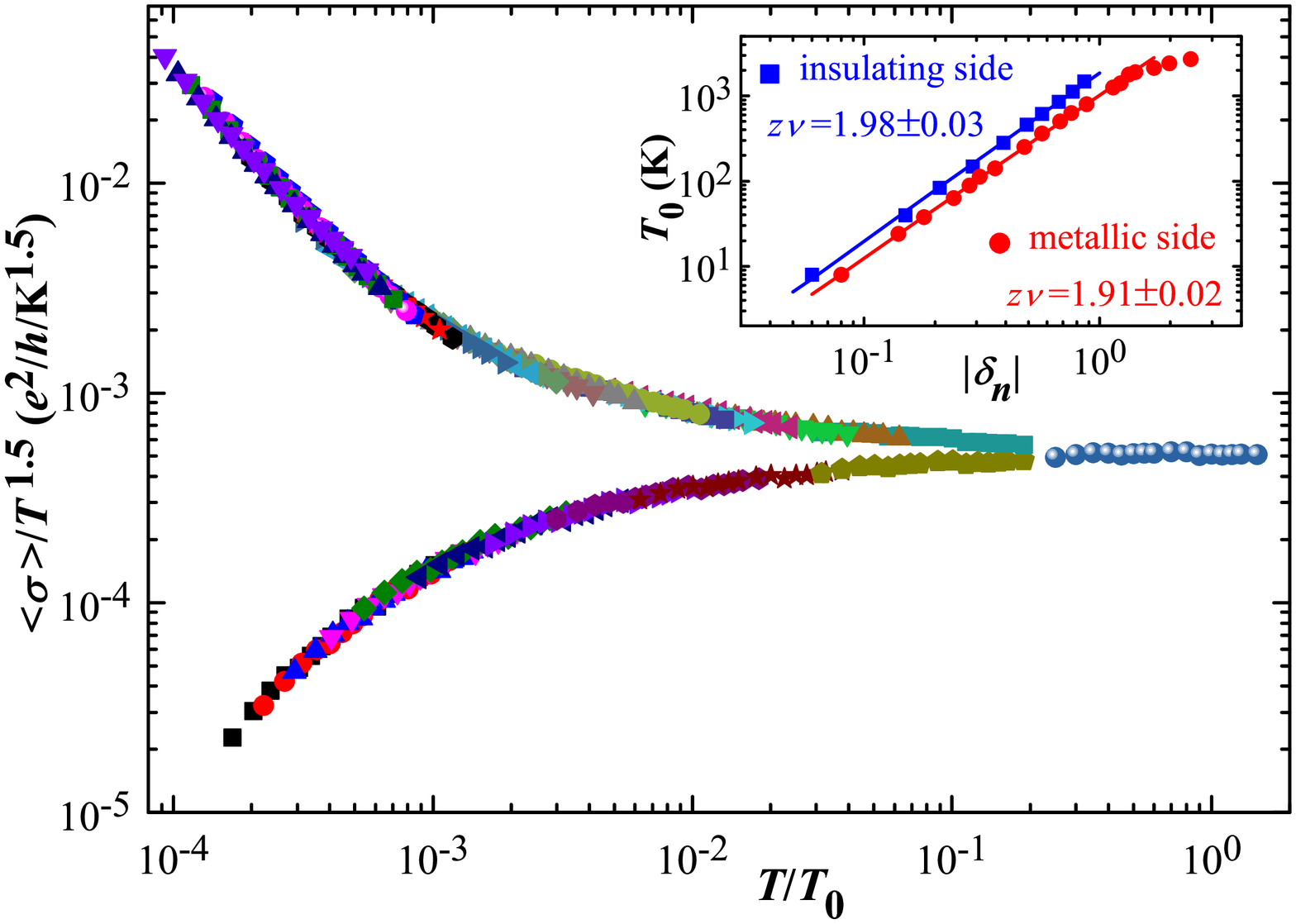}\llap{\parbox[b]{0.5in}{\textbf{(a)}\\\rule{0ex}{0.6in}}}
\includegraphics[width=9.3cm]{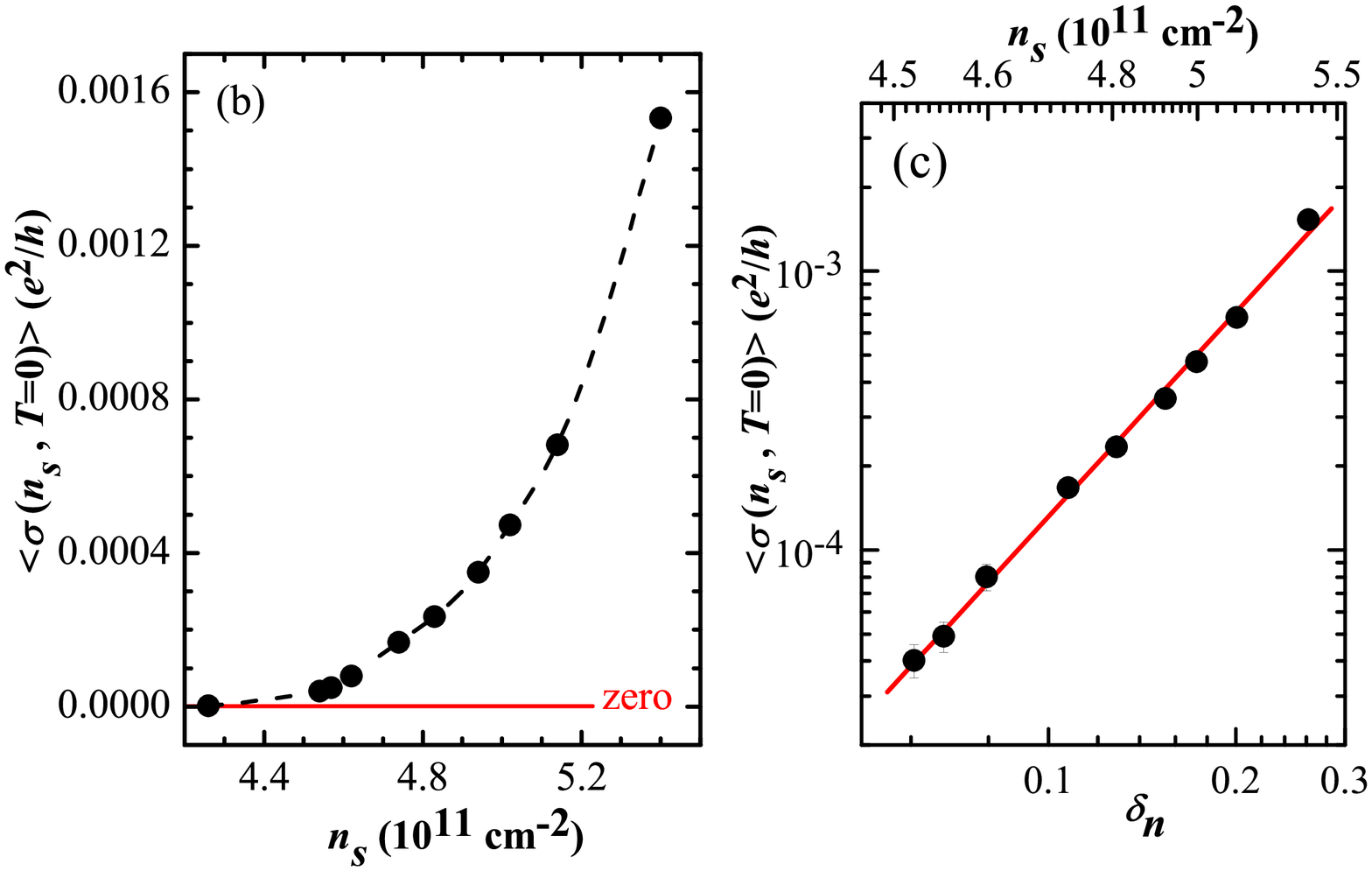}
\caption {(a) Scaling of $\sigma/\sigma_{c}\propto\sigma/T^{x}$, $x=1.5$, for a high-disorder sample with a thin oxide ($d_{ox}=6.9$~nm), i.e. with screened Coulomb interactions.  Different symbols correspond to $n_s$ from $3.40\times10^{11}$~cm$^{-2}$ to $6.70\times10^{11}$~cm$^{-2}$; $n_c= 4.26\times10^{11}$~cm$^2$.  It was possible to scale the data below about 1.5~K, over the range $0.007 \lesssim T/T_{F}\lesssim 0.04$.  Inset: $T_0$ \textit{vs.} $\delta_n$.  The lines are fits with slopes $z\nu=1.98\pm0.03$ and $z\nu=1.91\pm0.02$ on the insulating and metallic sides, respectively.  (b) $\sigma(n_s, T=0)$ \textit{vs} $n_s$.  The dashed line guides the eye.  (c) $\sigma(n_s, T=0)$ \textit{vs} $\delta_n=(n_s-n_c)/n_c$, the distance from the MIT.  The solid line is a fit with the slope equal to the critical exponent $\mu=2.7\pm 0.3$.  From \cite{Ping2015}.}\label{fig:scalingthin}
\end{figure}
%%%%%%%%%%%%%%%%%%%%%%%%%%%%%%%%%%%%%%%%%%%%%%%%%%%%%%%%%%%%%
%
obtained from the insulating side.  The power-law behavior $\sigma(n_s, T=0)\propto\delta_{n}^{\mu}$ [Fig.~\ref{fig:scalingthin}(c)] is in agreement with general expectations (\ref{eq:mu}).  The critical exponent $\mu=2.7\pm 0.3$.

\item Finally, in the vicinity of the MIT, the conductivity can be described by a scaling form (\ref{eq:scaling})  [Fig.~\ref{fig:scalingthin}(a)]; here $z\nu\approx2.0$.  As expected for a QPT, the same value of $z\nu$ is found, within experimental error, on both sides of the transition.  The consistency of the scaling analysis is confirmed by comparing $\mu=x(z\nu)= 3.0\pm0.3$ obtained from scaling with $\mu=2.7\pm 0.3$ found from the $T=0$ extrapolations of $\sigma(n_s,T=0)$.

\end{enumerate}

The most significant result is that the critical exponents in thin- and thick-oxide high-disorder devices are the same, and thus not sensitive to the range of the Coulomb interactions.  Indeed, in such a disorder-dominated MIT, it is plausible that the length scale of the Coulomb interactions does not seem to play a major role.  It is important to note, though, that there may be some other quantities that are more sensitive to the range of the Coulomb interactions and that would be affected by the proximity to the gate.  E.g. the fate of the glassy behavior in a 2DES with short-range Coulomb interactions remains an open question.

\subsection{Effects of magnetic field}
\label{sec:field}

Magnetic fields $B$ applied parallel to the 2DES plane couple only to electrons' spins and, therefore, they are often used to probe the importance of spin, as opposed to charge, degrees of freedom.  One of the main questions addressed in such studies in the context of the 2D MIT has been the fate of the metallic phase in a parallel $B$.  High-disorder 2DESs have not been investigated yet, but on low-disorder samples, it has been established that the metallic phase and the MIT survive in high parallel $B$ such that the 2DES is fully spin-polarized.  In particular, $(n_s, B, T=0)$ phase diagrams have been determined both in the absence \cite{Jan2004} and presence of local magnetic moments \cite{Eng2002}.  Just like in $B=0$, the properties of the MIT as a QPT have proved easier to study and analyze in systems with local moments, because of their stronger and better-defined $\sigma(T)$.  In the absence of local moments, the situation is more subtle, as described below.

\begin{figure}[t]
\centering
\includegraphics[width=9.4cm,clip]{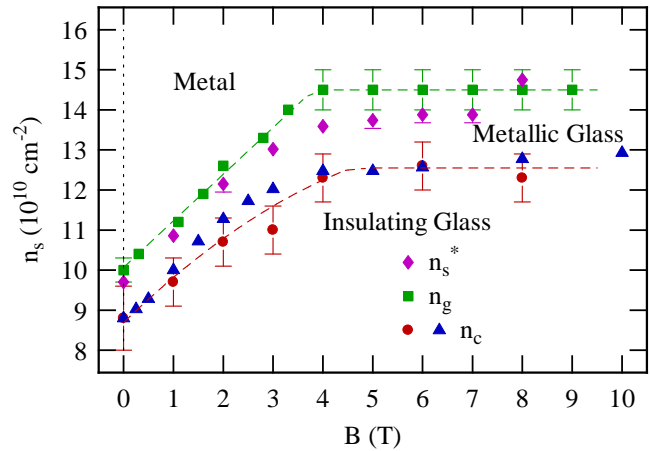}
\vspace*{-0.2in}
\caption { $T=0$ phase diagram for a low-disorder 2DES in a parallel magnetic field \cite{Jan2004}.  The dashed lines guide the eye.  The $n_c$ values are from \cite{Jan2004} (dots) and \cite{Shashkin-nc} (triangles).  The glass transition takes place at $n_g(B)>n_c(B)$, giving rise to an intermediate, metallic glass phase.  The density at the separatrix $n_{s}^{\ast}\approx n_g$ within the error for all $B$.}\label{fig:field}
\end{figure}

In zero field, studies of charge dynamics in a low-disorder 2DES (with no local moments) have established (Sec.~\ref{sec:lowdynamics}) that the onset of glassy behavior essentially coincides with the MIT, i.e. $n_g\approx n_c$.  A parallel $B$, however, gives rise to an intermediate, metallic glassy phase (Fig.~\ref{fig:field})  
%
%%%%%%%%%%%%%%%%%%%%%%
%%%%%%%%%%%%%%%%%%%%%%%%%%%%%%%%%%%%%%%%%%%%%%%%%%%%%%%%%%%%%
%
with the same, non-Fermi-liquid form $\sigma(n_s,B,T)=\sigma(n_s,B,T=0) +b(n_s,B)T^{1.5}$ as what was observed in high-disorder samples at $B=0$.  Once again, such a simple and well-defined $T$-dependence allows for reliable extrapolations to $T=0$, finding $n_c(B)$, $\sigma_c\propto T^{1.5}$, and $\sigma(n_s,B,T=0)\propto\delta_{n}^{\mu}$, $\mu\approx 1.5$, for a given $B$ \cite{Jan2004}, in agreement with theoretical expectations near a QPT.  Therefore, it is the emergence of the metallic glassy phase in $B\neq 0$ that makes it possible to determine $n_c$ also from the \verb"metallic" side of the MIT, the task that, in zero field, remains a challenge (Sec.~\ref{sec:low}).  The remarkable agreement between $n_c(B)$ obtained from $\sigma(T)$ on both insulating and metallic sides of the MIT are strong evidence for the survival of the MIT and the metallic phase in parallel $B$.  In contrast to highly disordered samples, here the intermediate phase spans a very narrow range of $n_s$ and, therefore, it can be observed only if $n_s$ is varied in fine steps.  Finally, detailed studies of the behavior in the metallic glassy phase and the MIT have been performed so far only up to about 4~T in Fig.~\ref{fig:field}, the field above which the 2DES is fully spin polarized.  Properties of the MIT at higher fields, that is, between a spin-polarized metal and a spin-polarized insulator, await future study.

In the presence of local magnetic moments, it is interesting that the $(n_s, B, T=0)$ phase diagram \cite{Eng2002} is quite similar to that shown in Fig.~\ref{fig:field}, even though the $d\sigma/dT$ behaviors in the metallic phase at $B=0$ are strikingly different (Sections \ref{sec:low} and \ref{sec:moments}).  This similarity probably results from the general expectation of a power-law shift of $n_c$ with $B$ in the case of a true MIT \cite{Belitz}.  Such a shift has been observed also in several 3D systems \cite{3Dbeta-1,3Dbeta-2,3Dbeta-3,3Dbeta-4}.

\subsection{Possible universality classes of the 2D metal-insulator transition}
\label{sec:universal}

All the critical exponents have been summarized in Table~\ref{table1} \cite{Ping2015}.  Based on their values, it appears that the 2D MIT in Si MOSFETs can be divided into three universality classes: a) high disorder, b) low disorder, and c) low disorder in the presence of scattering by local magnetic moments.  However, so far there have been no studies of high-disorder samples with local moments.

%
%%%%%%%%%%%%%%%%%% Table 1 %%%%%%%%%%%%%%%%%%%%%%%%%%
{\begin{table}[t]
\caption{Critical exponents $x$, $z\nu$, $\mu$, and calculated $\mu=x(z\nu)$ for 2D electron systems in Si MOSFETs with different disorder (from \cite{Ping2015}).  ``--'' indicates that the data are either insufficient or unavailable.  $\mu_{\textrm{peak}}$ is given in units of m$^2$/Vs, $d_{ox}$ in nm, and $n_c$ in $10^{11}$cm$^{-2}$.    \vspace{12pt}}
{\normalsize
\renewcommand\arraystretch{1.5}% adjust the table width
\centering
\newsavebox{\tableboxtwo}
\begin{lrbox}{\tableboxtwo}
\begin{threeparttable}
\begin{tabular}{|c|c|c|c|c|c|c|}
\hline 
 &\multicolumn{2}{c|}{High-disorder system}&\multicolumn{2}{c|}{\multirowcell{2}{Special disorder: local magnetic moments}}&\multicolumn{2}{c|}{\multirowcell{2}{Low-disorder\\ system}}\\ %[1ex] %adds vertical space
\cline{2-3}
 &\makecell{thin oxide}&\makecell{thick oxide}& \multicolumn{2}{c|}{} & \multicolumn{2}{c|}{}\\
    \hline
    $\mu_{peak}$  & 0.04 & 0.06 &\multicolumn{2}{c|}{$\sim 1$} & \multicolumn{2}{c|}{$\sim 1-3$}\\
    \hline
    $d_{ox}$ & 6.9 & 50 &\multicolumn{2}{c|}{43.5} &\multicolumn{2}{c|}{40-600} \\
    \hline
     & $B=0$ &$B=0$  & $B=0$  & $B\neq 0$ & $B=0$ & $B\neq 0$ \\
      \hline
    $n_c$  & $4.2\pm0.2$ & $5.0\pm0.3$ & 0.5-1 & \{$[n_c(B)/n_c(0)]-1$\} $\propto B$ & $\sim 1$ & \{$[n_c(B)/n_c(0)]-1$\} $\propto B$ \\
     \hline
    $x$ & $1.5\pm0.1$ & $1.5\pm0.1$ & $2.6\pm 0.4$ & $2.7\pm 0.4$  & -- & $1.5\pm 0.1$\\
    \hline
    $z\nu$ & $2.0\pm0.1$ & $2.1\pm0.1$ & $1.3\pm 0.1$ & $0.9\pm 0.3$ & $1.0-1.7$  & --\\
    \hline
    $\mu$ & $2.7\pm0.3$ & -- & $3.0\pm 0.1$ & $3.0\pm 0.1$ & 1-1.5 & $1.5\pm 0.1$\\
    \hline
    $\mu=x(z\nu)$ & $3.0\pm0.3$ & $3.3\pm0.4$ & $3.4\pm 0.4$ & $2.4\pm 1$ & -- & --\\
\hline
\end{tabular}
\end{threeparttable}
\end{lrbox}
\resizebox{\columnwidth}{!}{\usebox{\tableboxtwo}}   } \label{table1}
\end{table}
%%%%%%%%%%%%%%%%%%%%%%%%%%%%%%%%%%%%%%%%%%%%%%%%
%
Table~\ref{table1} also includes, where available, the values obtained in low parallel $B$ (\textit{i.e.} $B$ not high enough to fully spin polarize the 2DES \cite{polarization1, polarization2}).  In low parallel $B$, $[n_c(B)/n_c(0)-1]\propto B^{\beta}$  with $\beta=1.0\pm 0.1$ for low-disorder samples both in the absence \cite{Dolgopolov1992,Sakr,Shashkin-nc,Jan2004} and presence of scattering by local magnetic moments \cite{Eng2002}.  It is apparent that such low fields do not seem to affect any of the critical exponents.

On the other hand, there is a major difference between the values of $z\nu$ in high- and low-disorder devices, indicating that sufficiently strong disorder changes the nature of the MIT from the interaction-driven to disorder-dominated.  The possibility of a disorder-dominated 2D MIT has been demonstrated theoretically \cite{Punnoose} for both long-range and short-range interactions.  Although there is currently no microscopic theory that describes the detailed properties of the observed disorder-dominated MIT, it is interesting that in the available theories \cite{Belitz, Punnoose}, the range of the Coulomb interactions does not play a significant role, consistent with experimental observations.  In the interaction-driven MIT in low-disorder 2DESs (see also chapter by V. Dobrosavljevi\'c), the effect of the range of electron-electron interactions on the critical exponents still remains to be studied experimentally.

It should be also noted that percolation models \cite{percolation} cannot describe these findings. E.g. the 2D percolation $\mu\simeq 1.3$, as opposed to the much larger experimental $\mu\simeq 3$ in high-disorder samples and low-disorder devices with local magnetic moments (Table~\ref{table1}).  In fact, it is interesting that the same large $\mu\simeq 3$ is observed in those two types of samples, even though their values of $z\nu$ are very different ($z\nu\approx 2$ and $z\nu\approx 1.3$, respectively).  Therefore, while $z\nu$ seems to depend on the amount of disorder, the exponent $x$ instead appears to be more sensitive to the type of disorder (e.g. magnetic vs nonmagnetic).

In order to confirm the proposed universality classes of the 2D MIT (Table~\ref{table1}), which have been established based on studies of Si MOSFETs, it is clearly necessary to probe the behavior of 2D systems in other types of materials, in particular beyond conventional semiconductor heterostructures.  New families of 2D crystals, formed by extracting atomically thin layers of materials with weak interlayer van der Waals interactions \cite{Novoselov-review, 2Dcrystals-review}, represent specially promising candidates for such investigations.

\subsection{Metal-insulator transition in novel 2D materials}
\label{novel}

2D materials extracted from van der Waals solids, such as transition metal dichalcogenides (e.g. MoS$_2$, WS$_2$, MoSe$_2$, WSe$_2$), represent a new avenue for exploring quantum critical phenomena and the effects of dimensionality on correlated electronic phases, and may also lead to development of new electronic and optoelectronic applications.  Indeed, there is currently intensive activity in device fabrication based on 2D atomic layers, including FETs, photodetectors, light emitting devices, etc. \cite{TMD-devices}.  

Recently, FETs with mobilities as high as $\sim 0.03$~m$^{2}$/Vs at a few K have been reported in MoS$_2$  \cite{Kis-MoS2} and ReS$_2$ \cite{Luis-MIT}.  Those mobilities are comparable to $\mu_{\textrm{peak}}$ in high-disorder Si MOSFETs (Sec.~\ref{sec:high}), which had been fabricated using a commercial $0.25$-$\mu$m Si technology \cite{Taur}.  In both MoS$_2$ and ReS$_2$ FETs, metallic $d\sigma/dT <0$ behavior was observed at high $n_s\gtrsim 10^{13}$cm$^{-2}$, and the transition to $d\sigma/dT>0$ temperature dependence at somewhat lower $n_s$ was attributed to a 2D MIT.  Although all the material parameters may not be known precisely \cite{Luis-MIT}, it is estimated that $r_s\sim 4$ near the apparent MIT in both types of devices, i.e. a bit smaller than in high-disorder 2DES in Si where $r_s\sim 7$.   

It is interesting that, in ReS$_2$, $\sigma(T)$ exhibits nonmonotonic behavior at high $n_s$ [Fig.~\ref{fig:Luis}(a)] reminiscent of
%
%%%%%%%%%%%%%%%%%%%%%%
\begin{figure}[t]
\centering
\includegraphics[width=5.5cm,clip]{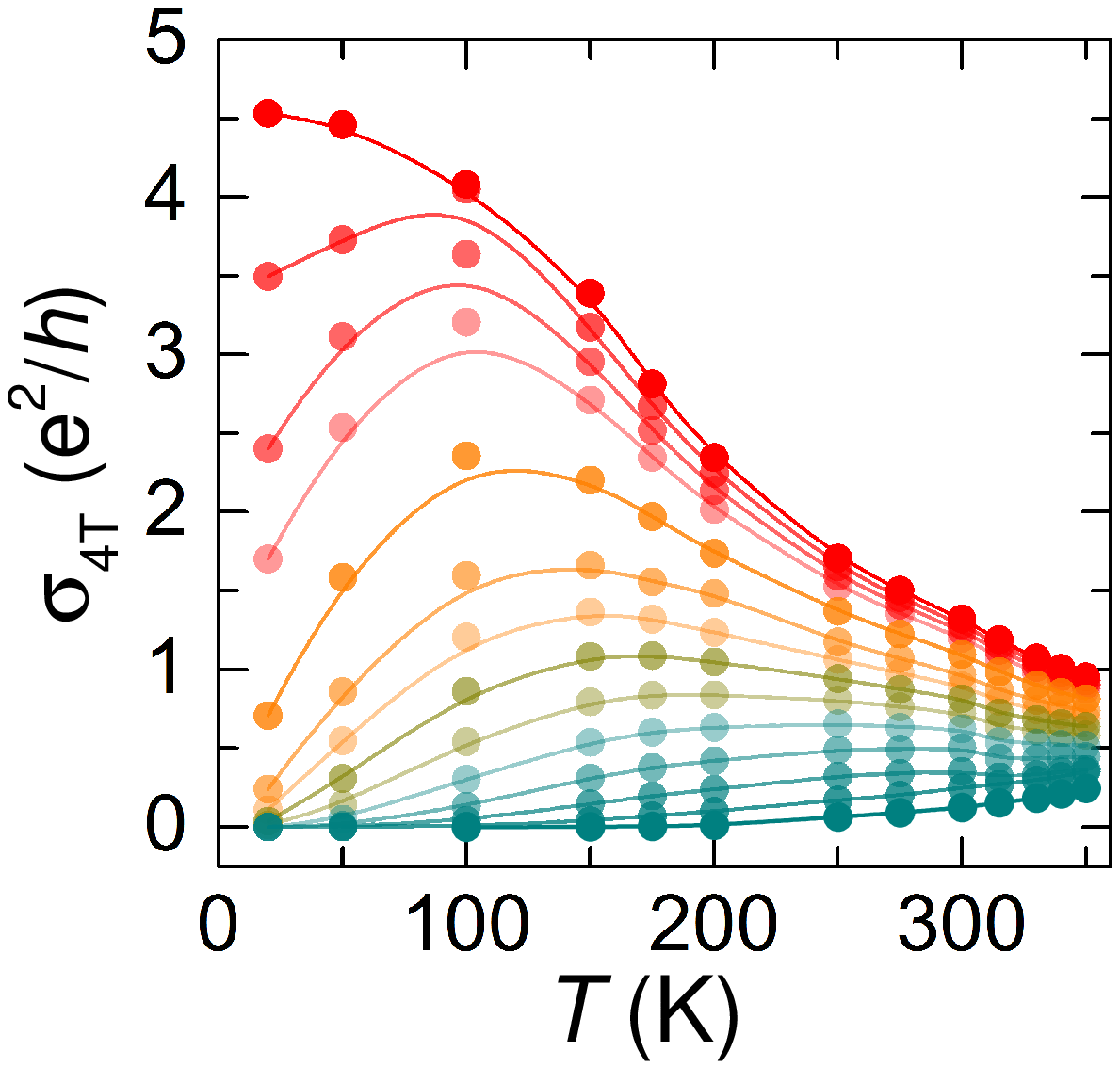}\llap{\parbox[b]{0.4in}{\textbf{(a)}\\\rule{0ex}{1.7in}}}
\includegraphics[width=7.3cm,clip]{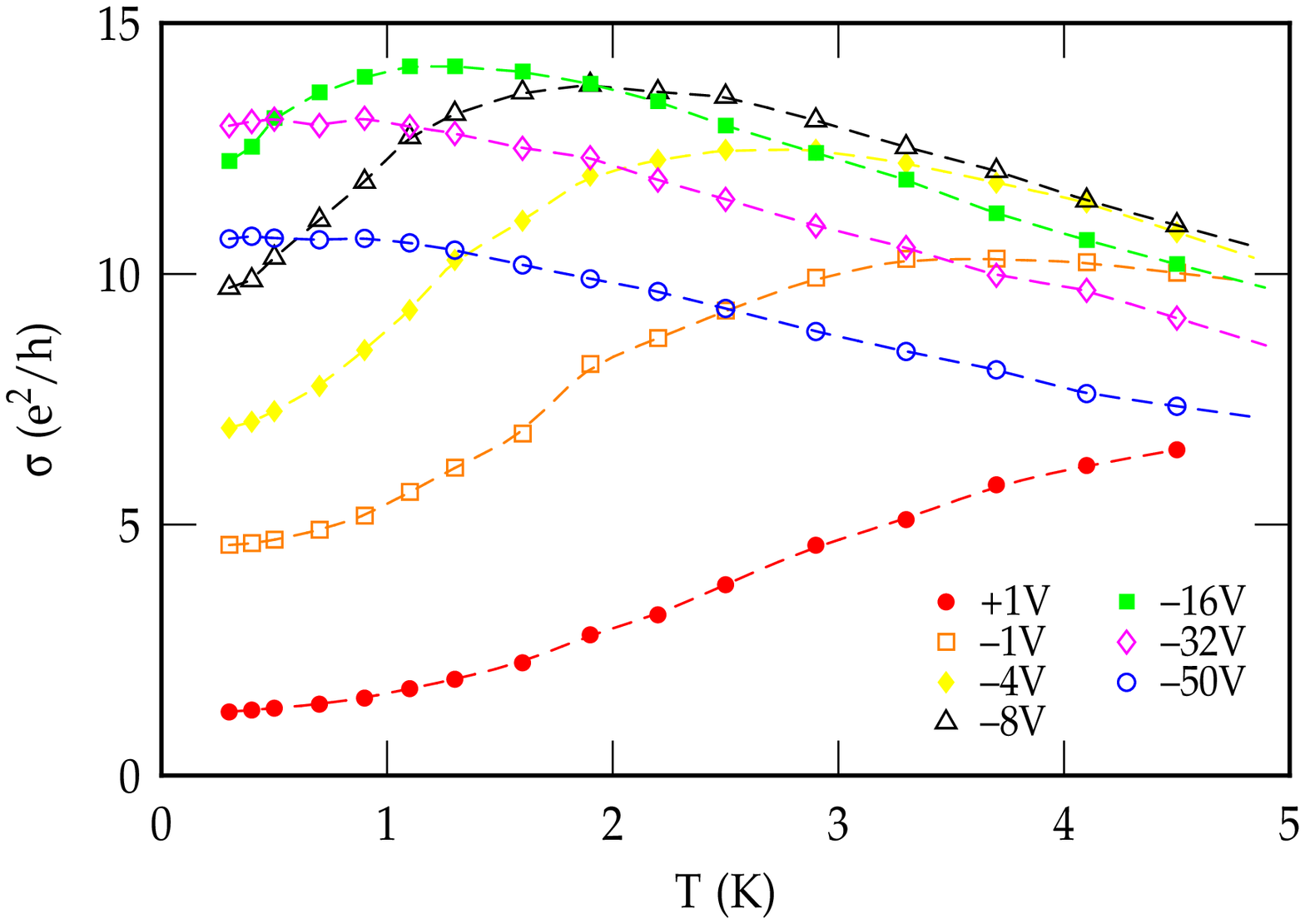}\llap{\parbox[b]{0.4in}{\textbf{(b)}\\\rule{0ex}{1.7in}}}
\caption {(a) Four-terminal conductivity of a few-layer ReS$_2$ vs $T$ for different $n_s$, which are controlled by the substrate (back-gate) bias (from \cite{Luis-MIT}.  (b) $\sigma$ vs $T$ in a low-disorder Si MOSFET for a fixed $n_s=3.0\times 10^{11}$cm$^{-2}$ and different $V_{sub}$, as shown (from \cite{Feng1999}.}\label{fig:Luis}
\end{figure}
%%%%%%%%%%%%%%%%%%%%%%%%%%%%%%%%%%%%%%%%%%%%%%%%%%%%%%%%%%%%%
%
a low-disorder 2DES in which scattering by local magnetic moments is controlled by varying the substrate bias.  Figure~\ref{fig:Luis}(b), for example, shows the effect of $V_{sub}$ in a Si MOSFET for a fixed $n_s>n_c$: while $d\sigma/dT<0$ for a large negative $V_{sub}=-50$~V, reducing the negative $V_{sub}$ leads to the emergence of a maximum in $\sigma(T)$, such that $d\sigma/dT>0$ in the entire experimental $T$ range for $V_{sub}=+1$~V.  It is understood \cite{Feng1999,Feng2001} that, for $T$ below the maximum, the scattering is dominated by local magnetic moments (Sec.~\ref{sec:moments}; also Fig.~\ref{fig:sigmaT}).  Most notably, it has been shown \cite{Luis-MIT} that $\sigma(T)$ of few-layer ReS$_2$ FETs at temperatures below the maximum obey dynamical scaling (\ref{eq:scaling}) with the exponents $x\approx 2.3$, $z\nu\approx 1.3$, and $\mu\approx 2.9$, the latter being consistent with $\mu=x(z\nu)\approx 3.0$.  It is most striking that these values indeed agree, within error, with the exponents established for (low-disorder) samples in which local moments dominate (Table~\ref{table1}).\footnote{As noted in Sec.~\ref{sec:universal}, there have been no studies of high-disorder samples with local magnetic moments.}

Although the agreement between critical exponents obtained on ReS$_2$ and those on 2DES in Si is encouraging, there are several caveats.  For example, in ReS$_2$ there is some uncertainty in the determination of $n_s$ as a function of $V_{sub}$ so that, depending on the method used, the data analysis may yield a different set of exponents (e.g. $x\approx 3.4$, $z\nu\approx 0.6$, $\mu\approx 2.1$; \cite{Luis-MIT}).  In addition, scaling was performed at fairly high $T\lesssim T_{F}$ and over a limited range of $n_s$ in the insulating regime.  Therefore, not only do the measurements need to be extended to much lower $T$ and $n_s$, but also it is important to gain better understanding of the basic FET and material characteristics before reliable results on the 2D MIT can be obtained on ReS$_2$ and other novel 2D materials.

\section{Charge Dynamics Near the 2D Metal-Insulator Transition and the Nature of the Insulating State}
\label{sec:dynamics}

Section~\ref{sec:critical} focused on the critical region near the 2D MIT, in particular on describing conductivity measurements that have demonstrated dynamical scaling (\ref{eq:scaling}), the main signature of the MIT as a quantum phase transition.  The critical exponents, which have been determined for both interaction-driven and disorder-dominated MIT (Sec.~\ref{sec:universal}), represent a property of the quantum critical point.  However, these studies do not provide information about the nature of the metallic and insulating phases.  The chapter by A. A. Shashkin and S. V. Kravchenko discusses various experimental results obtained in the \verb"metallic" regime of low-disorder 2DES.  In contrast, this Section focuses on experiments that probe the nature of the \verb"insulating" state, as well as charge dynamics \verb"across the MIT", in both high- and low-disorder 2DESs.

There is substantial evidence that in many materials near the MIT, both strong electronic correlations and disorder play an important role, and thus their competition is expected to lead to glassy behavior of electrons, in analogy with other frustrated systems \cite{Mir-Dob-review, Miranda-Vlad-CIQPT}.  A common denominator for all glasses is the existence of a complex or ``rugged'' free energy landscape, consisting of a large number of metastable states, separated by barriers of different heights.  This results in phenomena such as slow, nonexponential relaxations, divergence of the equilibration time, and breaking of ergodicity, i.e. the inability of the system to equilibrate on experimental time scales.  Therefore, such out-of-equilibrium systems also exhibit aging effects \cite{aging1,glasses}, where the response to an external excitation (i.e. relaxation) depends on the system history in addition to the time $t$.  A detailed analysis of temporal fluctuations (noise) of the relevant observables yields complementary information on configurational rearrangements or transitions between metastable states.\footnote{In equilibrium systems, the connection between spontaneous fluctuations of a variable and the response of such a variable to a small perturbation in its conjugated field is given by the fluctuation-dissipation relation.  See \cite{nieuw-book} for the review and discussion of thermodynamics of out-of-equilibrium systems.}  Non-Gaussian distributions of various observables in glassy systems have been reported \cite{hetero-review}, reflecting the presence of large, \textit{collective} rearrangements.  Therefore, the two basic ways to probe the dynamics of glassy systems involve studies of relaxations and fluctuations.

Most experimental studies of charge or Coulomb glasses have focused on situations where electrons are strongly localized due to disorder, i.e. deep in the insulating regime and far from the MIT \cite{ArielAmir-review}.  In recent years, however, studies of both relaxations and fluctuations (or noise) in Si MOSFETs have provided evidence for out-of-equilibrium or glassy dynamics of the 2DES in the insulating regime, near the MIT, and just on the metallic side of the transition in the intermediate, metallic glassy phase (Fig.~\ref{fig:summary}).  Those results, described below (see also \cite{Dragana-CIQPT}), impose strong constraints on the theories for the 2D MIT, and should be also helpful in understanding the complex behavior near the MIT in a variety of strongly correlated materials.

\subsection{High-disorder 2D electron systems}
\label{sec:highdynamics}

Glassy charge dynamics in a high-disorder 2DES was probed using several different experimental protocols, all of which can be divided into two groups: one of them involves applying a large (with respect to $E_F$) perturbation to the system and studying the relaxations of conductivity, and the other one involves a study of conductivity fluctuations with time as a result of a small perturbation.  In all experiments, $k_{B}T$ was the lowest energy scale.

A \verb"large perturbation" was applied to the 2DES by making a large charge in $V_g$ or carrier density, i.e. such that $k_{B}T\ll E_{F} < \Delta E_{F}$.  In one protocol, which involved a study of the relaxations $\sigma(t)$ following a rapid change of $n_s$, the following key manifestations of glassiness were established \cite{relax-PRL} for all $n_s$ below the glass transition density $n_g$ ($n_c<n_g$, i.e. $n_g$ is on the metallic side of the MIT; Fig.~\ref{fig:summary}).

\begin{enumerate}

\item The temperature dependence of the equilibration time $\tau_{\textrm{eq}}$ obeys a simply activated form, so that $\tau_{\textrm{eq}}\rightarrow \infty$ as $T\rightarrow 0$.  The diverging equilibration time means that, strictly speaking, the system cannot reach equilibrium only at $T=0$, i.e. the glass transition temperature $T_g=0$.

\item At low enough $T$, however, $\tau_{\textrm{eq}}$ can easily exceed experimental times (e.g. at 1~K, $\tau_{\textrm{eq}}$ is estimated to exceed the age of the Universe by several orders of magnitude!), so that the system appears glassy: for $t<\tau_{\textrm{eq}}$, the relaxations obey a nonexponential form, which reflects the existence of a broad distribution of relaxation times.

\item Nonexponential relaxations obey dynamical scaling $\sigma(t,T)/\sigma_{0}(T)\propto t^{-\alpha(n_s)}\exp[-(t/\tau(n_s, T))^{\beta(n_s)}]$, where $0<\alpha(n_s)<0.4$, $0.2<\beta(n_s)<0.45$, such that in the $T\rightarrow 0$ limit, the relaxations attain a pure power-law form $\sigma/\sigma_0 \propto t^{-\alpha}$.   The dynamical scaling and the power-law relaxation at $T_g$ are consistent with the general scaling arguments \cite{scaling} near a continuous phase transition occurring at $T_g=0$, similar to the discussion in Sec.~\ref{sec:intro}.

\end{enumerate}

A key characteristic of relaxing glassy systems is the loss of time translation invariance, reflected in aging effects \cite{aging1,aging2,glasses}.  Therefore, in another protocol, relaxations $\sigma(t)$ were studied after a temporary change of $n_s$ during the waiting time $t_\textrm{w}$.  The sample history was varied by changing $t_\textrm{w}$ and $T$ for several initial (final) $n_s$.  The main results include the following.

\begin{enumerate}

\item It was demonstrated that the 2DES exhibits aging, and the conditions that lead to memory loss and nonmonotonic response were identified precisely \cite{tw-PRL}.

\item There is an abrupt change in the nature of the glassy phase exactly at the 2D MIT itself, before glassiness disappears completely at a higher density $n_g$:  (a) while the so-called full aging\footnote{In case of full or simple aging, the aging function $\sigma(t,t_\textrm{w})$ exhibits scaling with $t/t_\textrm{w}$.} is observed in the insulating regime ($n_s<n_c$), there are significant departures from full aging in the metallic glassy phase, i.e for $n_c<n_s<n_g$; (b) the amplitude of the relaxations peaks just below the MIT, and it is strongly suppressed in the insulating phase \cite{DP-aging, aging-PhysB}.

\item As the system ages and slowly approaches equilibrium, the non-Gaussian conductance noise becomes increasingly Gaussian \cite{Ping2012}, similar to the behavior of a great variety of out-of-equilibrium systems.

\end{enumerate}

The results of the aging studies represent strong evidence that the insulating glassy phase and the metallic glassy phase are different.   It should be also noted that, in the mean-field models of glasses, for example, two different cases are distinguished: one, where full aging is expected, and the other, where no $t/t_w$ scaling is expected \cite{mean-field}.  Therefore, the difference in the aging properties below and above $n_c$ puts constraints on the theories of glassy freezing and its role in the physics of the 2D MIT.

In the second group of experiments, a \verb"small perturbation" was applied either by making a small change in $n_s$, such that $k_{B}T<\Delta E_{F}\ll E_{F}$ \cite{Snezana2002, SB_PhysE, JJ_PRL02, Ping2012}, or by cooling, i.e. making a small change in $T$, such that $k_{B}T<k_{B}\Delta T\ll E_{F}$ \cite{Ping2012}.  In both cases, there were no observable relaxations of $\sigma$, but large non-Gaussian conductance noise emerged for $n_s<n_g$.  The non-Gaussian nature of the noise indicates that the fluctuating units are correlated.  

The noise was studied by analyzing the full probability distribution of the fluctuations (or the probability density function, PDF), power spectrum, and the so-called second spectrum, which is a fourth-order noise statistic.  The analysis indicates that the slow, correlated behavior is consistent with the so-called hierarchical picture of glassy dynamics, similar to conventional, metallic spin glasses \cite{Binder}.  In that scenario, the system wanders collectively between many metastable states related by a kinetic hierarchy.  Metastable states correspond to the local minima or ``valleys'' in the free energy landscape, separated by barriers with a wide, hierarchical distribution of heights and, thus, relaxation times. Intervalley transitions, which are reconfigurations of a large number of electrons, thus lead to the observed strong, correlated, $1/f$-type noise.

Therefore, the 2DES has many characteristics in common with a large class of both 2D and 3D out-of-equilibrium systems, strongly suggesting that many such universal features are robust manifestations of glassiness, regardless of the dimensionality of the system. However, it appears that there are also some effects that may be unique to Coulomb glasses.  This is illustrated in Fig.~\ref{fig:PDF}, for example, where PDFs obtained after cooling (left column) are 
%
%%%%%%%%%%%%%%%%%%%%%%
\begin{figure}[t]
\centering
\includegraphics[width=8.5cm,clip]{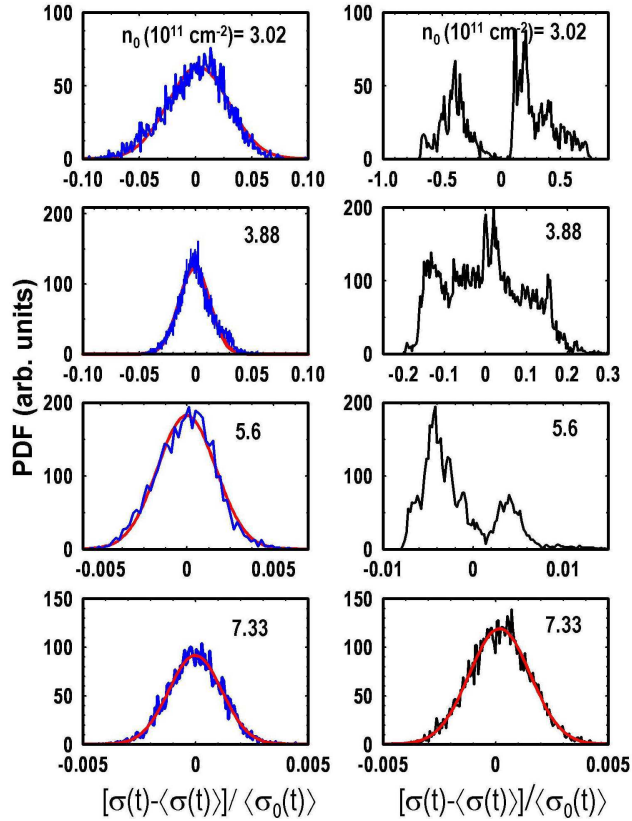}
\caption {History dependence of the probability density functions of the fluctuations measured in a high-disorder 2DES at $T=0.24$~K for several carrier densities $n_0 (10^{11}$cm$^{-2}$), as shown; $n_g\approx 7.5\times 10^{11}$cm$^{-2}$ (from \cite{Ping2012}).  $\langle\sigma_0\rangle$ is the time-averaged conductivity corresponding to $n_0$ at the measurement $T$, obtained after cooling.  The red curves are fits to a Gaussian distribution.  Left column: PDFs of the noise after cooling from 10~K to 0.24~K for each given $n_0$.  The cooling time was at least one hour long. In this protocol, $\langle\sigma(t)\rangle\equiv\langle\sigma_0\rangle$, i.e. there are no observable relaxations after cooling.  Right column: PDFs of the noise measured after a subsequent change of $n_s$ from $n_0$ to a much higher value $n_1=20.26\times 10^{11}$cm$^{-2}$ during $t_w=1000$~s.  In this protocol, $\langle\sigma(t)\rangle$ describes the slowly relaxing background.}\label{fig:PDF}
\end{figure}
%%%%%%%%%%%%%%%%%%%%%%%%%%%%%%%%%%%%%%%%%%%%%%%%%%%%%%%%%%%%%
%
compared to those obtained after a subsequent large change of $n_s$ during the waiting time $t_\textrm{w}$ (right column).  What is striking is that, even though the PDFs were measured under exactly the same experimental conditions, they look remarkably different for all $n_s<n_g$.  In particular, the non-Gaussian PDFs obtained after cooling are smooth, single-peaked functions, reminiscent of PDFs in a variety of systems displaying critical \cite{criticalPDF1,criticalPDF2}, glassy \cite{hetero-review}, or other out-of-equilibrium behavior (e.g. the Danube water level \cite{Danube}).  In all these systems, the PDFs are skewed, resembling a zero-centered Gaussian, which describes pseudoequilibrium fluctuations, with one (exponential) tail that is due to large, rare events.  In contrast, a temporary change in $n_s$ results in complex, multipeaked, random-looking PDFs.  Similar complicated, multipeaked PDFs were observed also after a small change of $n_s$ (not shown).  Therefore, these results demonstrate not only that noise depends on history, as may be expected in a glassy system, but also that the change of $n_s$ has a qualitatively different and more dramatic effect than $\Delta T$.  In fact, the results strongly suggest that the density change reshuffles all energies, because of the Coulomb interactions, thus modifying the free energy landscape of the 2DES.  For this reason, theoretical modeling of the glassy dynamics in this system might be considerably more difficult than in some other types of glassy materials.

\subsection{Low-disorder 2D electron systems}
\label{sec:lowdynamics}

Studies of charge dynamics in low-disorder 2DESs have so far included only \verb"small perturbation" protocols, i.e. measurements of the conductance fluctuations following a small change of $n_s$, including the analysis of the higher-order statistics \cite{JJ_PRL02}.  Qualitatively, the same behavior was observed as in high-disorder samples: there is a well-defined density $n_g$ below which the noise becomes non-Gaussian and increases by several orders of magnitude as $n_s$ or $T$ are reduced.  This slow, correlated noise is consistent with the hierarchical pictures of glassy dynamics, as described above.  The only difference is that, in a low-disorder 2DES, the intermediate glassy phase practically vanishes and the glass transition coincides with the MIT ($n_g\approx n_c$).

Noise measurements were performed also in parallel $B$ \cite{Jan2004}.  By adopting the same criteria for the glass transition as in zero field, it was possible to determine $n_g(B)$ shown in Fig.~\ref{fig:field}, identify the emergence of the intermediate, metallic glassy phase (see Sec.~\ref{sec:field}), and to establish that charge, not spin, degrees of freedom are responsible for glassy ordering.  Therefore, the results demonstrate that the 2D MIT is closely related to the melting of this Coulomb glass.  

In fact, experiments on both high- and low-disorder 2DESs in Si strongly support theoretical proposals describing the 2D MIT as the melting of a Coulomb glass \cite{Darko-glass,MIT-glassothers1,MIT-glassothers2,MIT-glassothers3,Vlad-MITglass1,Dalidovich}.  In particular, a model with Coulomb interactions and sufficient disorder \cite{Darko-glass} predicts the emergence of an intermediate metallic glass phase.  The theoretical work, however, still needs to be  extended to studies of the critical behavior, including critical exponents.    Experimentally, it would be interesting to perform relaxation studies on low-disorder samples to look for aging phenomena, which may also provide further insights into the nature of the insulating, Coulomb glass phase.

\section{Conductor-Insulator Transition and Charge Dynamics in Quasi-2D Strongly Correlated Systems}

Many novel materials are created by doping an insulating host and thus are close to a conductor-insulator transition.  For example, there is now broad agreement \cite{Lee-highTc} that the problem of high-temperature superconductivity in copper oxides is synonymous to that of doping of a Mott insulator.   Arguments have been put forward that, in weakly doped Mott insulators near the MIT, the system will settle for a nanoscale phase separation between a conductor and an insulator \cite{phasesep1, phasesep2, phasesep3}.  This leads to the possibility for a myriad of competing charge configurations and the emergence of the associated glassy dynamics, perhaps even in the absence of disorder \cite{Schmalian}.  Indeed, cuprates and many other materials exhibit various complex phenomena due to the existence of several competing ground states \cite{Elbio}, thus providing strong impetus towards the better understanding of the MIT and the behavior of the charge degrees of freedom.  However, experimental studies near the conductor-insulator transition in many materials, such as cuprates, are complicated by the accompanying changes in magnetic or structural symmetry.  On the other hand, those materials are usually layered, with weak interlayer coupling, so that, in most instances, they behave effectively as 2D systems.  Therefore, comparative studies of the MIT and charge dynamics in 2DESs in semiconductor heterostructures and in strongly correlated quasi-2D materials, such as cuprates, present an especially promising approach in addressing the problem of complexity near the MIT in strongly correlated systems.  Such studies should make it possible to separate out effects that are more universal from those that are material specific. 

In La$_{2-x}$Sr$_x$CuO$_4$ (LSCO), the prototypical cuprate high-temperature superconductor, conductance (or resistance) noise spectroscopy was employed at very low $T$ \cite{IR_PRL, IR-inplane} to probe charge dynamics in the lightly doped regime where the ground state is insulating.  Here the charge carriers, doped holes, seem to populate areas that separate the hole-poor antiferromagnetic (AF) domains located in CuO$_2$ planes.  The magnetic moments in different domains are known to undergo cooperative freezing at a temperature $T_{\textrm{SG}}$ ($\sim$ a few K) into a ``cluster spin glass'' phase.  The noise measurements demonstrated the emergence of slow, correlated dynamics and nonergodic behavior at very low $T\ll T_{\textrm{SG}}$, deep inside the spin-glass phase, which rules out spins as the origin of the observed glassy dynamics.  In addition, all the noise characteristics were found to be insensitive to both the magnetic field and the magnetic history, further indicating that the observed glassiness reflects the dynamics of charge, not spins. This is analogous to the magnetic insensitivity of the noise in a glassy, fully spin-polarized 2DES in Si \cite{Jan2004}.  The gradual enhancement of the glassy behavior in LSCO with decreasing $T$ strongly suggests that the phase transition to a charge glass state occurs at $T_g=0$, similar to a 2DES (Sec.~\ref{sec:dynamics}).  In contrast to a 2DES, however, the non-Gaussian statistics in LSCO is not consistent with the hierarchical picture of glasses, but rather reflects the presence of some characteristic length scale.  This result supports the picture of spatial segregation of holes into interacting, hole-rich droplets or clusters, which are separated by hole-poor AF domains.

Additional evidence for charge-glass behavior at low $T$ in lightly doped, insulating LSCO was found from dielectric \cite{Glenton} and magnetotransport \cite{Ivana-MR} studies.  The observed history-dependent resistance and hysteretic magnetoresistance \cite{IR_PRL, Ivana-MR, Ivana-stripes} were then used to explore the key question in the physics of cuprates, namely how such an insulating, dynamically heterogeneous ground state evolves with doping and gives way to high-temperature superconductivity \cite{Xiaoyan-memory, Xiaoyan-NM}.  The $(x, T, H)$ phase diagram (Fig.~\ref{fig:LSCO}), where $x$ (in this Section) is doping, $H$ is a
%
%%%%%%%%%%%%%%%%%%%%%%
\begin{figure}
\centering
\includegraphics[width=9.0cm,clip]{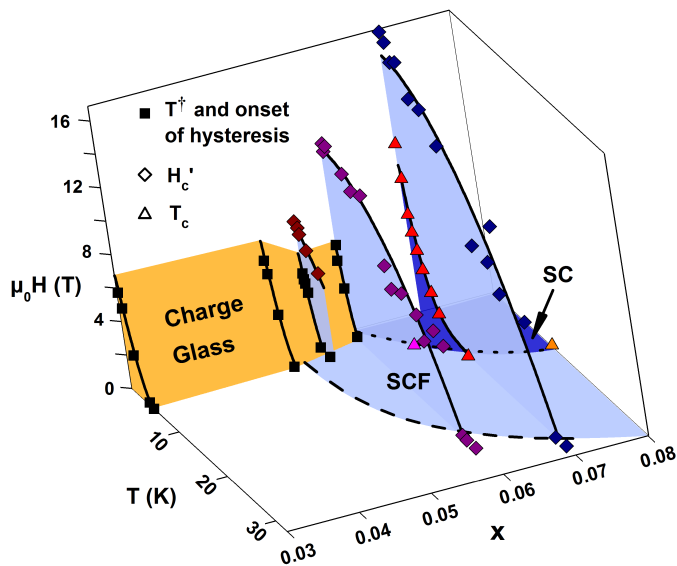}
\caption {Phase diagram shows the evolution of the glassy region and the emergence of superconducting fluctuations (SCFs) and superconductivity (SC) with doping, temperature and magnetic field in La$_{2-x}$Sr$_x$CuO$_4$ (from \cite{Xiaoyan-NM}).  The extent of the glassy regime does not depend on the field orientation. The range of SCFs is shown for the field applied perpendicular to CuO$_2$ planes.  Solid and dashed lines guide the eye.  Different colors of symbols for $H_{c}'(T)$ and $T_c(H)$ correspond to different values of doping.}\label{fig:LSCO}
\end{figure}
%%%%%%%%%%%%%%%%%%%%%%%%%%%%%%%%%%%%%%%%%%%%%%%%%%%%%%%%%%%%%
%
magnetic field perpendicular to CuO$_2$ planes, shows that a collective, glassy state of charge clusters located in CuO$_2$ planes is suppressed by increasing the doping.  At the same time, adding charge carriers leads to the formation of localized Cooper pairs (or superconducting fluctuations) already within this insulating, intrinsically heterogeneous charge-ordered state, consistent with the so-called Bose glass picture for the transition from an insulator to a superconductor (SIT) \cite{MPAF-SITRc}.  Surprisingly, it was also found (not shown) that the superconducting fluctuations on the insulating side were quenched at low temperatures by the charge glass order. Therefore, the pair localization and the onset of SIT in LSCO are influenced by a competing charge order, and not merely by disorder, as seems to be the case in some conventional superconductors.  Those observations provide a new perspective on the mechanism for the SIT.

The experiments discussed above were carried out on samples in which the number of carriers was varied by chemical doping.  In order to study scaling associated with the $T=0$ superconductor-insulator transition, however, that method has the disadvantage that the carrier concentration cannot be tuned continuously, and also it alters the level of disorder in the material.  Therefore, electrostatic charging, similar to that in semiconductor heterostructures, would be preferable. Recently, it has become possible to electrostatically induce large concentration changes in a variety of novel materials \cite{egating-Ahn-review}.  In cuprates, ionic liquids have been used to make FET-like devices to study scaling near the SIT in thin films of hole-doped La$_{2-x}$Sr$_x$CuO$_4$ \cite{Ivan-scaling}, YBa$_2$Cu$_3$O$_{7-x}$ \cite{Goldman-YBCO-scaling} and La$_2$CuO$_{4+\delta}$ \cite{Goldman-LCO-scaling}, and electron-doped Pr$_{2-x}$Ce$_x$CuO$_4$ \cite{electrondoped-scaling}.  The observed scaling appeared consistent with the so-called bosonic picture \cite{MPAF-SITRc}, in which Cooper pairs are localized on the insulating side of the SIT and the critical resistivity is independent of $T$.\footnote{The additional presence of some fermionic excitations at finite $T$ on the insulating side was suggested for the electron-doped cuprate~\cite{electrondoped-scaling}.}  The obtained critical exponents $z\nu$ were $1.5, 2.2, 1.2,$ and 2.4, respectively, but the reasons for that difference are not understood.  Moreover, in contrast to the 2DES in Si (Sec.~\ref{sec:critical}), the scaling was done over a limited range of parameters: typically, $T/T_0$ spans only about an order of magnitude, and the lowest $T\sim$ a few K.  Clearly, it would be important to extend these studies to much lower $T$.  Interestingly, it was found that scaling in YBa$_2$Cu$_3$O$_{7-x}$ breaks down below 6~K, suggesting that the SIT may not be direct and may involve an intermediate phase \cite{Goldman-YBCO-scaling}.  Unfortunately, there were not sufficient data available to resolve this issue (the lowest $T$ in that experiment was 2~K), so that the questions about the nature of the carrier-concentration-driven SIT and whether it involves an intermediate phase in the $T\rightarrow 0$ limit remain open for all cuprates.  

Finally, in analogy to a 2DES in Si (Sec.~\ref{sec:dynamics}), electrostatic doping opens up an exciting possibility to investigate glassy relaxations that may accompany quantum phase transitions in cuprates and other strongly correlated systems.  Such studies should provide a much better understanding of the phases and quantum criticality.

\section{Conclusions}

Experimental studies of the critical behavior of conductivity in a variety of two-dimensional electron systems in Si provide strong evidence for the existence of a sharp, $T=0$ metal-insulator transition at low electron densities regardless of the amount of disorder.  The critical exponents obtained from dynamical scaling suggest that there are several universality classes of the 2D MIT, depending on the amount and type of disorder.  Since other types of experiments carried out on the metallic side of the transition indicate that the MIT in low-disorder samples is driven by electron-electron interactions, this implies that the MIT in high-disorder samples is dominated by disorder.  In both cases, however, Coulomb interactions between electrons must play a key role in stabilizing the metallic phase.  

The role of long-range Coulomb interactions is further revealed in the studies of charge dynamics across the MIT, which demonstrate that, in both low- and high-disorder systems, the insulating state is a Coulomb glass.  The peculiarity of a 2DES with a high amount of disorder is the emergence of an intermediate phase between the metal and the insulator, which is poorly metallic and glassy.  The aging properties demonstrate, however, that the nature of the metallic glassy phase is different from that of the insulating Coulomb glass.

While the above experimental findings await theoretical description and understanding, novel 2D materials, such as those extracted from van der Waals solids, as well as quasi-2D strongly correlated materials, including cuprates, present promising new avenues for testing the generality of the observed phenomena and gaining further insight into the problem of the MIT.

\section*{Acknowledgements}

This work was supported by NSF Grant No. DMR-1307075 and the National High Magnetic Field Laboratory through NSF Cooperative Agreement No. DMR-1157490 and the State of Florida.

\end{document}